\documentclass[prx,twocolumn,superscriptaddress,notitlepage]{revtex4-2}

\usepackage{graphicx}
\usepackage{amsmath,amsfonts,amsthm,amssymb} 
\usepackage{times}
\usepackage{physics} 
\usepackage[dvipsnames]{xcolor}
\usepackage{hyperref}
\hypersetup{
    colorlinks=true,
    linkcolor=RoyalBlue,
    citecolor=RoyalBlue,
    filecolor=black,      
    urlcolor=black,
    pdftitle={Overleaf Example},
    pdfpagemode=FullScreen,
    }

\renewcommand{\figurename}{Fig.} 

\begin{document}

\title{Squeezing Classical Antiferromagnets into Quantum Spin Liquids via Global Cavity Fluctuations}

\author{Charlie-Ray Mann}
\affiliation{ICFO--Institut de Ciencies Fotoniques, The Barcelona Institute of Science and Technology, 08860 Castelldefels (Barcelona), Spain.}

\author{Mark A. Oehlgrien}
\affiliation{ICFO--Institut de Ciencies Fotoniques, The Barcelona Institute of Science and Technology, 08860 Castelldefels (Barcelona), Spain.}

\author{Błażej Jaworowski}
\affiliation{ICFO--Institut de Ciencies Fotoniques, The Barcelona Institute of Science and Technology, 08860 Castelldefels (Barcelona), Spain.}

\author{Giuseppe Calajó}
\affiliation{Dipartimento di Fisica e Astronomia “G. Galilei”, via Marzolo 8, I-35131 Padova, Italy.}
\affiliation{Istituto Nazionale di Fisica Nucleare (INFN), Sezione di Padova, I-35131 Padova, Italy.}

\author{Jamir Marino}
\affiliation{Institut für Physik, Johannes Gutenberg Universität Mainz D-55099 Mainz, Germany.}
\affiliation{Department of Physics, The State University of New York at Buﬀalo, Buﬀalo, New York 14260, USA.}

\author{Kyung S. Choi}
\affiliation{Q-Block Computing Inc., Edmonton, Alberta, T5J 3S9, Canada.}

\author{Darrick E. Chang}
\affiliation{ICFO--Institut de Ciencies Fotoniques, The Barcelona Institute of Science and Technology, 08860 Castelldefels (Barcelona), Spain.}
\affiliation{ICREA--Instituci\'o Catalana de Recerca i Estudis Avan\c{c}ats, 08015 Barcelona, Spain.}

\begin{abstract}  
\noindent Cavity quantum electrodynamics with atomic ensembles is typically associated with collective spin phenomena, such as superradiance and spin squeezing, in which the atoms evolve collectively as a macroscopic spin ($S \sim N/2$) on the Bloch sphere. Surprisingly, we show that the tendency toward a collective spin description need not imply collective spin phenomena; rather, it can be exploited to generate new forms of strongly correlated quantum matter. The key idea is to use uniform cavity-mediated interactions to energetically project the system into the total-spin singlet sector ($S=0$)---a highly entangled subspace where the physics is governed entirely by cavity fluctuations. Focusing on Rydberg atom arrays coupled to a single-mode cavity, we show that global cavity fluctuations can effectively squeeze classical antiferromagnets into quantum spin liquids, characterized by non-local entanglement, fractionalized excitations, and emergent gauge fields. This work suggests that cavity QED can be a surprising resource for inducing strongly correlated phenomena, which could be explored in the new generation of hybrid tweezer-cavity platforms.
\end{abstract}

\maketitle

%%%%%%%%%%%%%%%%%%%%%%%%%%%%%%%%%%%%%

\vspace{2mm}
\noindent \textbf{\textit{Introduction}}.--- A hallmark of cavity quantum electrodynamics (QED) is its ability to mediate uniform all-to-all interactions between atoms via a delocalized cavity mode. This feature has been used to prepare collectively entangled spin-squeezed states \cite{leroux2010implementation, bohnet2014reduced, hosten2016measurement, cox2016deterministic, pedrozo2020entanglement, cooper2024graph, robinson2024direct}---a key resource for quantum metrology---and to simulate non-equilibrium phenomena such as superradiance \cite{baumann2010dicke, bohnet2012steady, norcia2016superradiance, song2025dissipation}, time crystals \cite{kessler2021observation, kongkhambut2022observation}, and dynamical phase transitions \cite{muniz2020exploring, young2024observing}. Yet the same all-to-all connectivity typically limits the many-body complexity by confining the dynamics to the manifolds of macroscopic collective spin ($S \sim N/2$). As a result, mean-field or Gaussian theories often provide an accurate description---and can even become exact in the thermodynamic limit---restricting access to strongly correlated phenomena that are now actively explored in other synthetic quantum platforms \cite{semeghini2021probing, satzinger2021realizing, iqbal2024non, xu2024non, evered2025probing, will2025probing}.

To unlock access to richer many-body phenomena in cavity QED, recent efforts have focused on engineering spatial structure into cavity-mediated interactions through two complementary approaches. The first strategy is to modify the atomic degrees of freedom: for instance, using magnetic-field gradients and multi-tone drives to spatially modulate the atom–light coupling and program arbitrary interaction graphs \cite{periwal2021programmable}. The second strategy is to modify the cavity architecture to enrich the mode structure: examples include multimode confocal cavities that generate frustrated interaction landscapes that support spin-glass behavior \cite{marsh2024entanglement,kroeze2025directly}, and twisted ring cavities that imprint synthetic magnetic fields on the photons \cite{schine2016synthetic,clark2020observation}.

However, we show that uniform all-to-all interactions can in fact be a natural resource for generating strongly correlated phenomena. The central idea is to use strong-cavity-mediated interactions to energetically project the system into the total-spin singlet sector ($S=0$)---a highly entangled subspace where the physics is governed entirely by quantum fluctuations of the cavity field, thereby explicitly evading the collective spin paradigm. Crucially, this singlet sector is exponentially degenerate, and even a weak spatially structured perturbation is sufficient to lift the degeneracy and stabilize a strongly correlated singlet as the true ground state.

This cavity projection paradigm can be realized in emerging tweezer-cavity platforms  \cite{kong2021melting,deist2022mid,liu2023realization,grinkemeyer2025error,hartung2024quantum,hu2025site}, where programmable short-range Rydberg interactions can be engineered in flexible lattice geometries. Motivated by this novel setting, we consider a classical Ising model coupled to a single-mode cavity. While projecting into entangled subspaces is generally challenging, within the singlet sector we establish an exact mapping to a short-range Heisenberg model---revealing an emergent notion of locality despite the presence of strong all-to-all interactions. This mapping provides a guiding framework for engineering a diverse landscape of correlated singlet states, including various quantum spin liquids (QSLs). These exotic phases lie beyond Landau’s symmetry-breaking paradigm and are instead characterized by non-local entanglement, fractionalized excitations, and emergent gauge fields \cite{savary2016quantum}.

For concreteness, we focus on a family of antiferromagnetic $J_1$–$J_2$ Ising models and map the ground-state phase diagram. On the square and triangular lattices, weak global cavity fluctuations initially squeeze the classical antiferromagnet (AFM) ground states of the Ising Hamiltonian, generating EPR-like entanglement between sublattices. Beyond a critical coupling strength, these cavity fluctuations can destabilize the long-range magnetic order, melting the squeezed AFMs into QSLs. In contrast, on the kagome lattice, even a weak cavity perturbation is sufficient to lift the extensive classical degeneracy and stabilize a QSL ground state.  These results challenge the prevailing view that all-to-all interactions inevitably lead to collective spin physics, while raising fundamental questions about the nature of strongly correlated phenomena in nonlocal quantum systems.

%------------ FIGURE 1 -------------% 
\begin{figure}[t]
\centering
\includegraphics[width=0.45\textwidth]{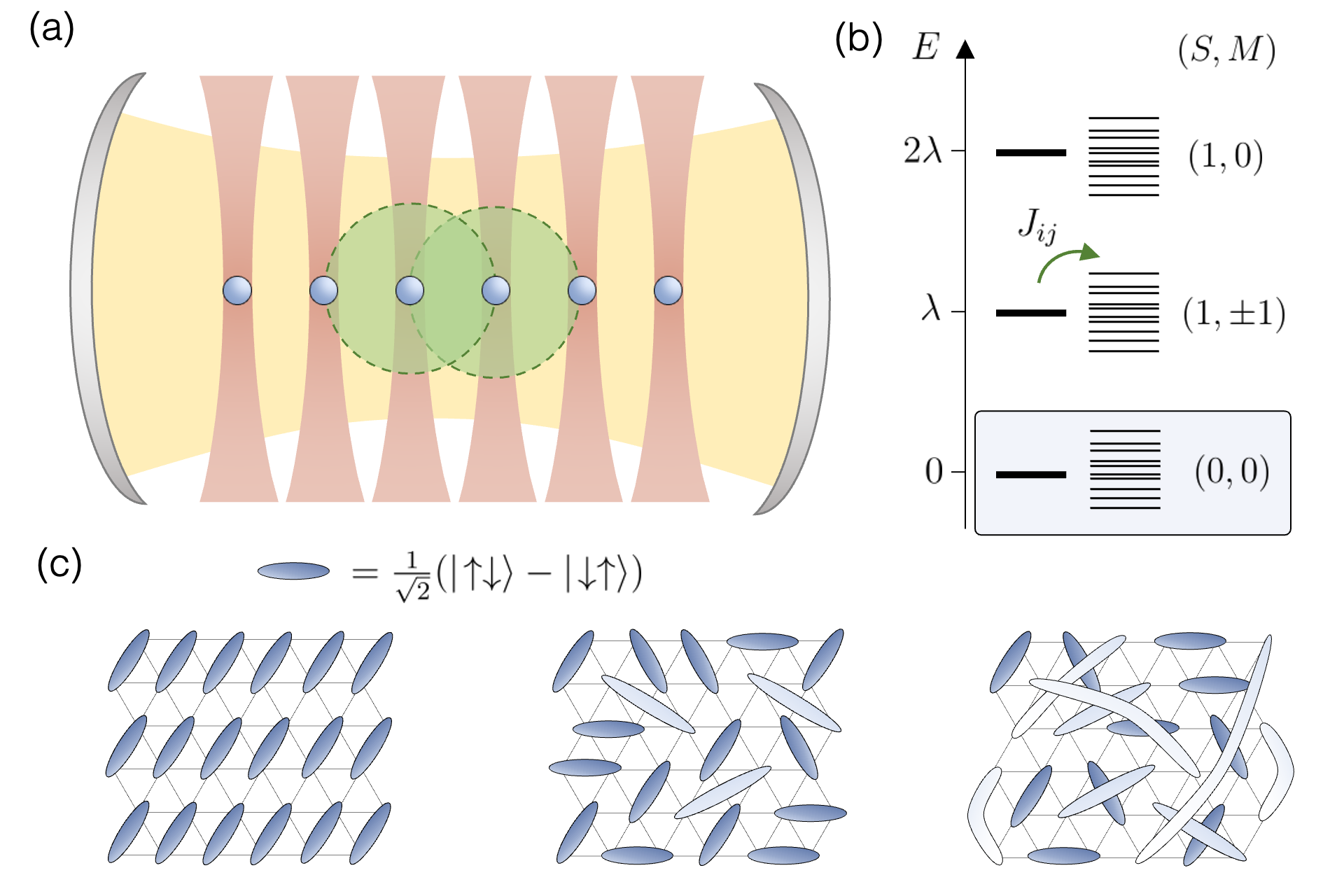}
\caption{\textbf{Cavity projection into the singlet sector}. (a) Schematic of a hybrid tweezer-cavity platform. A delocalized cavity mode (yellow) mediates uniform all-to-all interactions, while Rydberg interactions (green circles) generates competing short-range Ising couplings on a programmable tweezer array. (b) Structure of the many-body spectrum of the TCI model in the strong-cavity limit, which decouples into blocks labeled by total-spin quantum numbers $(S, M)$, with the lowest energy block corresponding to the singlet sector ($0,0$). (c) Examples of singlet coverings within the exponentially large degenerate manifold of ground states of the all-to-all cavity Hamiltonian ($J_{ij}=0$), composed of pairwise spin singlets (blue ellipses) of arbitrary spatial range.}
\label{fig:Singlet_Projection}
\end{figure}
%------------ FIGURE 1 -------------%

\vspace{1mm}
\noindent \textbf{\textit{Tavis-Cummings-Ising spin model}}.--- We consider a spin-$1/2$ model on an arbitrary lattice with $N$  sites described by the generalized XXZ Hamiltonian
%--------------------- 
\begin{equation} 
    H = \lambda (S_x^2 + S_y^2) + \sum_{i<j} J_{ij} S_z^i S_z^j\,, 
\label{eq:TCI_Model} 
\end{equation} 
%----------------------
which we refer to as the Tavis-Cummings-Ising (TCI) spin model. Here, $J_{ij}$ encodes the pattern of short-range Ising couplings, and $S_\alpha^i$ are spin-$1/2$ operators at lattice site $i$ with $\alpha\in \{x,y,z\}$. Moreover, $\lambda$ encodes the strength of the all-to-all spin-exchange interactions, which can be expressed in terms of collective spin operators $S_\alpha = \sum_i S_\alpha^i$.  

This target Hamiltonian can be realized in several platforms, such as tweezer-cavity interfaces (Fig.\,\ref{fig:Singlet_Projection}\textcolor{RoyalBlue}{a}). As a concrete implementation scheme, the qubit can be encoded in two hyperfine ground states with a tunable spin-cavity interaction engineered via cavity-assisted Raman transitions \cite{dimer2007proposed}, while tunable short-range interactions can be generated using Rydberg dressing techniques \cite{zeiher2016many}. In the regime of large detuning from the excited atomic states and cavity photons, both can be adiabatically eliminated, yielding the effective TCI spin model in Eq.\,\eqref{eq:TCI_Model}.

Although similar cavity-Ising models have been studied previously \cite{lee2004first,gammelmark2011phase,gammelmark2012interacting,zhang2013rydberg,zhang2014quantum,gelhausen2016quantum,rohn2020ising,schuler2020vacua,an2022quantum,puel2024confined,bacciconi2025local}, these works have primarily focused on parameter regimes that realize a superradiant phase in the strong-cavity limit, where the cavity field acquires a macroscopic expectation value and mean-field theory provides a reliable description. In the spin model picture, this corresponds to ferromagnetic interactions ($\lambda < 0$), where the ground state of the cavity Hamiltonian lies in the maximum total-spin sector ($S = N/2$). By contrast, we focus on the antiferromagnetic regime ($\lambda > 0$), where the physics is governed entirely by quantum fluctuations of the cavity field. Consequently, standard mean-field descriptions fail to capture the essential physics, necessitating a fundamentally different approach. 

In this direction, a recent theoretical study proposed that QSLs could arise in an SU(2)-symmetric Heisenberg model with competing short-range and power-law interactions generated by a multimode cavity \cite{chiocchetta2021cavity}. While suggestive, they employed a parton mean-field construction, which, although useful for identifying candidate phases, lacks predictive power as it is not variational without projection into the physical spin subspace. In contrast, we avoid such constructions entirely, but instead provide critical insight by establishing a map to well-studied models in condensed matter physics.

\vspace{1mm}
\noindent \textbf{\textit{Heisenberg singlet mapping}}.--- First, we will elucidate the nature of the ground state in the strong-cavity limit $\lambda/J_{ij}\to \infty $. In this regime, the energy spectrum effectively decouples into blocks corresponding to the total-spin quantum numbers $(S,M)$, where $0\leq S\leq N/2$ and $|M|\leq S$ (Fig.\,\ref{fig:Singlet_Projection}\textcolor{RoyalBlue}{b}). To leading order in Schrieffer–Wolff perturbation theory, the effective Hamiltonian is given by the projection of the Ising Hamiltonian into each total-spin sector
%----------------------
\begin{equation}
    H\simeq E_{S,M}+\sum_{S,M}\mathbb{P}_{S,M}H_{\mathrm{Ising}}\mathbb{P}_{S,M}\,,
    \label{eq:Projected_Hamiltonian}
\end{equation}
%----------------------
where $\mathbb{P}_{S, M}$ denotes the projector into the $(S, M)$ sector and $E_{S,M}=\lambda[S(S+1)-M^2]$ is the spectrum of the cavity Hamiltonian. Note that some total-spin sectors are degenerate, but they are not connected by the Ising interaction.

Interestingly, the lowest energy manifold corresponds to the singlet sector ($S=0$). In the absence of Ising interactions ($J_{ij}=0$), this sector is exponentially degenerate, spanned by all possible singlet coverings on the lattice (Fig.\,\ref{fig:Singlet_Projection}\textcolor{RoyalBlue}{c}). Physically, this degeneracy arises from the permutational invariance of the cavity Hamiltonian, which lacks any notion of dimensionality or lattice structure---any two spins can form singlets, regardless of their separation distance. Introducing short-range Ising interactions ($J_{ij}\neq0$) breaks this permutational symmetry, which lifts the degeneracy and selects a unique singlet ground state, which will generically be a valence bond solid (VBS) or a resonating valence bond state. 

Although projecting into highly entangled subspaces is generally challenging, the special structure of the singlet manifold offers a powerful simplification. Specifically, all singlet states are invariant under global SU(2) spin rotations, satisfying $R(\Omega)\ket{S=0}=\ket{S=0}$, where $R(\Omega)=\mathrm{e}^{-\mathrm{i}\psi S_z}\mathrm{e}^{-\mathrm{i}\varphi S_y}\mathrm{e}^{-\mathrm{i}\phi S_z}$ is a global spin rotation operator parameterized by Euler angles $\Omega=(\psi,\varphi,\phi)$. The projector onto the sector subspace can thus be expressed as a uniform average over the SU(2) group
%----------------------
\begin{equation}
     \mathbb{P}_{0,0}=\int \frac{\mathrm{d}\Omega}{8\pi^2}  R(\Omega)\,.
     \label{eq:Singlet_Projector}
\end{equation}
%----------------------
Consequently, any rotated Hamiltonian $R^\dagger(\Omega)H_{\mathrm{Ising}}R(\Omega)$ yields the same projected effective Hamiltonian---the cavity effectively enforces rotational invariance. This means we can replace the physical Ising Hamiltonian with its SU(2)-symmetrized form
%----------------------
\begin{equation}
     H_{\mathrm{Heis}}=\int \frac{\mathrm{d}\Omega}{8\pi^2}  R^\dagger(\Omega)H_{\mathrm{Ising}}R(\Omega)=\frac{1}{3}\sum_{i<j} J_{ij}\mathbf{S}_i \cdot \mathbf{S}_j\,,
     \label{eq:Heisenberg_Model}
\end{equation}
%----------------------
which is just the corresponding short-range Heisenberg model, where the exchange interactions $J_{ij}$ are inherited from the Ising couplings. This implies that, in the strong-cavity limit, the low-energy sector of the TCI model becomes equivalent to the singlet sector of the short-range Heisenberg model---realizing an emergent notion of locality, despite the dominance of the cavity-mediated all-to-all interactions. 

%------------ FIGURE 2 -------------% 
\begin{figure}[t]
\centering
\includegraphics[width=0.48\textwidth]{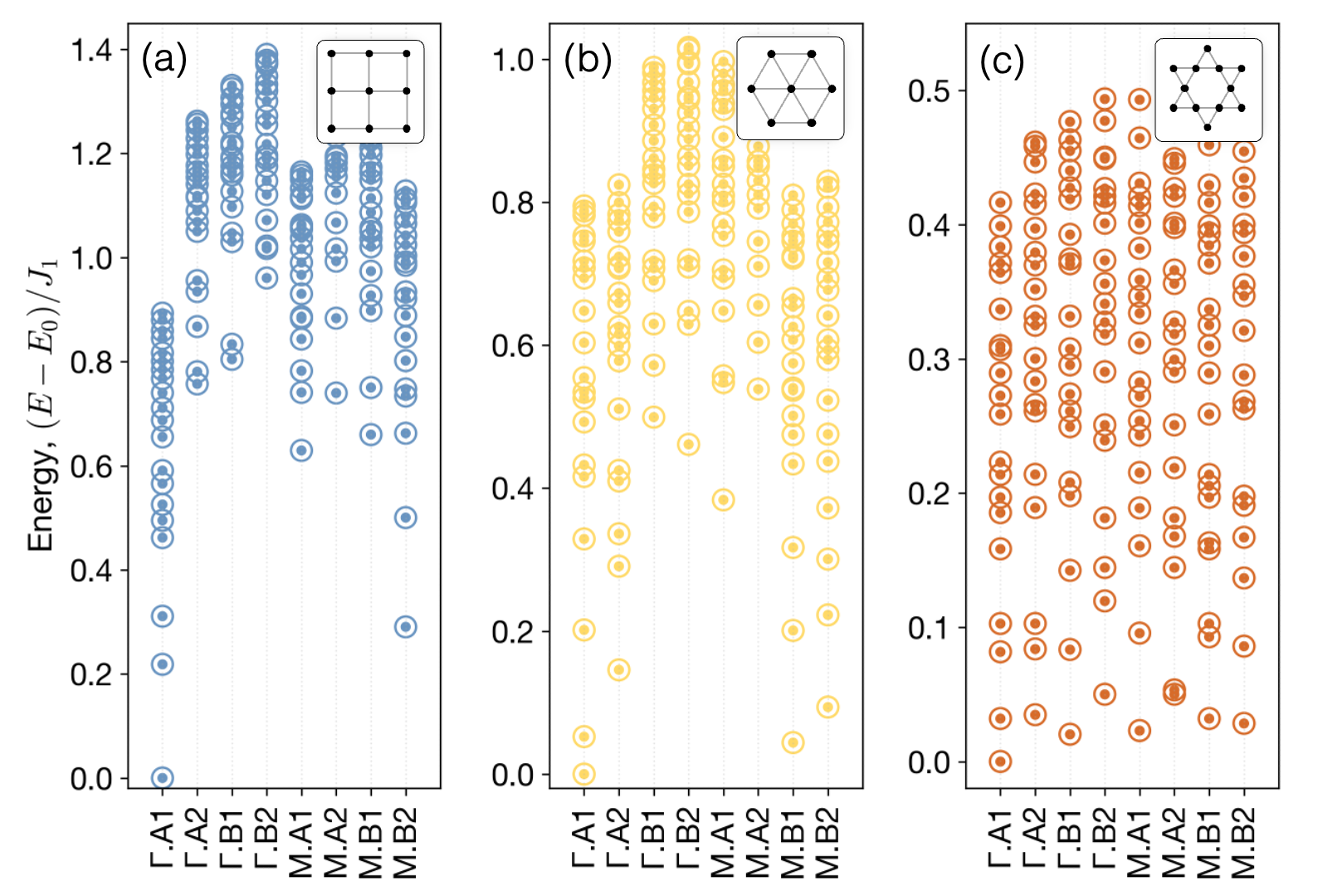}
\caption{\textbf{Heisenberg singlet mapping.} Low-energy spectrum of the $J_1$-$J_2$ TCI model (filled circles) on the (a) square lattice ($J_2/J_1 = 0.5$), (b) triangular lattice ($J_2/J_1 = 0.125$), and (c) kagome lattice ($J_2/J_1 = 0$) in the strong-cavity regime ($\lambda/J_1 = 10^3$). Each system contains $N = 24$ spins with periodic boundary conditions, and only the lowest 20 eigenvalues are shown within a few representative symmetry sectors (see Methods). These spectra are compared with the singlet sector of the corresponding $J_1$-$J_2$ Heisenberg models (open circles), which are predicted to host QSLs. In all cases, we observed a clear one-to-one correspondence between all eigenstates.}
\label{fig:Heisenberg_Mapping}
\end{figure}
%------------ FIGURE 2 -------------%

This Heisenberg mapping is particularly insightful when the Ising couplings are antiferromagnetic ($J_{ij}>0$). The Marshall-Lieb–Mattis theorem \cite{marshall1955antiferromagnetism,lieb1962ordering} guarantees that, on any finite bipartite lattice, the ground state of an antiferromagnetic Heisenberg model resides in the total-spin singlet sector ($S=0$). In practice, a singlet ground state is also found for a broad class of non-bipartite (frustrated) models. In such cases, the singlet ground states of the two models should be equivalent in the strong-cavity limit, allowing one to leverage the deep understanding of Heisenberg models to predict and design strongly correlated singlet states in this cavity QED setting. By contrast, for ferromagnetic couplings ($J_{ij}<0$), the singlet manifold corresponds to a high-energy sector of the Heisenberg model, which remains largely unexplored.

Here we focus on realizing QSLs, and thus consider a family of antiferromagnetic $J_1$–$J_2$ TCI models with nearest ($J_1$) and next-nearest ($J_2$) neighbor Ising couplings. Their short-range Heisenberg counterparts have long served as paradigmatic models in frustrated quantum magnetism and have been central to the modern understanding of QSLs. Extensive numerical studies---using exact diagonalization (ED), density matrix renormalization group (DMRG), two-dimensional tensor-network methods, variational projected parton wavefunctions, and machine-learning approaches---have identified nonmagnetic regions widely believed to host QSL (and/or VBS) phases on the square lattice ($0.45\lesssim  J_2/J_1\lesssim 0.6$) \cite{richter2010spin,jiang2012spin,hu2013direct,wang2013constructing, wang2018critical, nomura2021dirac,liu2022gapless}, triangular lattice ($0.07\lesssim  J_2/J_1\lesssim 0.15$)  \cite{zhu2015spin,hu2015competing,iqbal2016spin,hu2019dirac,wietek2024quantum}, and kagome lattice ($0\lesssim  J_2/J_1\lesssim 0.2$)  \cite{yan2011spin,jiang2012identifying,depenbrock2012nature,kolley2015phase,iqbal2013gapless,liao2017gapless,he2017signatures}. The precise nature of these QSLs remains the subject of intense debate, but leading candidates include both gapped and gapless $Z_2$ topological QSLs \cite{jiang2012spin, hu2013direct,wang2013constructing,zhu2015spin,hu2015competing,
yan2011spin,jiang2012identifying,depenbrock2012nature}, as well as gapless $U(1)$ Dirac QSLs \cite{nomura2021dirac, iqbal2016spin,hu2019dirac,wietek2024quantum,
iqbal2013gapless,liao2017gapless,he2017signatures}.

We certainly do not attempt to contribute to this debate here, but simply emphasize that the TCI model should inherit many key features of the QSLs in the strong-cavity limit. For example, in a gapped $Z_2$ QSL the singlet sector is expected to host the topologically degenerate ground states and vison excitations, while in a $U(1)$ Dirac QSL the singlet sector hosts the ground state, emergent photons and monopole excitations. To validate the mapping, we computed the energy spectra of the TCI and Heisenberg models on the square, triangular, and kagome lattices using ED on 24-site clusters (see Methods). As expected, in all cases we find a one-to-one correspondence between the low-energy eigenstates of the TCI model and the singlet sector of the short-range Heisenberg model (Fig.\,\ref{fig:Heisenberg_Mapping}\textcolor{RoyalBlue}{a-c}). 

However, the Heisenberg mapping does not extend trivially to other total-spin sectors $(S\neq 0)$, since one cannot invoke the same symmetrizing principle. As a result, the fate of the fractionalized spinon excitations remains an open question even if their properties were, in principle, well understood in the short-range Heisenberg model. In particular, the TCI and Heisenberg models possess different global spin-rotation symmetries, implying distinct projective symmetry group classifications \cite{reuther2014classification}, and the spinons may acquire qualitatively new dynamical properties in the presence of non-local couplings.

Outside the nonmagnetic regions, the Heisenberg singlet ground states exhibit different patterns of long-range magnetic order. These states are characterized by the static spin structure factor $\mathcal{S}_{\alpha\alpha}(\mathbf{q})=N^{-1}\sum_{ij}\mathrm{e}^{\mathrm{i} \mathbf{q}\cdot(\mathbf{r}_i-\mathbf{r}_j)} \langle S_\alpha^i S_\alpha^j\rangle$, which develops an extensive peak at the ordering wavevector $\mathbf{Q}$, located at different high-symmetry points within the Brillouin zone (see Methods). On the square lattice, we expect Néel AFM order ($\mathbf{Q}=\mathrm{M}$) for $J_2/J_1 \lesssim 0.45$ and stripe AFM order ($\mathbf{Q}=\mathrm{X}$) for $J_2/J_1 \gtrsim 0.6$. On the triangular lattice, we expect $120^\circ$ AFM order ($\mathbf{Q}=\mathrm{K}$) for  $J_2/J_1\lesssim  0.07$ and stripe AFM order ($\mathbf{Q}=\mathrm{M}$) for $J_2/J_1\gtrsim  0.15$, while on the kagome lattice we expect $120^\circ$ AFM order ($\mathbf{Q}=0$) for $J_2/J_1\gtrsim  0.2$.

Another key distinction between the short-range Heisenberg and TCI models concerns spontaneous symmetry breaking. In the Heisenberg case, the long-range magnetic order is accompanied by the so-called tower of states: a sequence of low-lying excitations in higher total-spin sectors ($S \neq 0$) whose energy splittings vanish as $\sim 1/N$. In the thermodynamic limit, coherent superpositions within this tower generate symmetry-broken states with finite order parameter $\langle S_\alpha^i\rangle\neq 0$. In the TCI model, this mechanism is fundamentally altered: the cavity term imposes a large energy penalty on higher-spin sectors, preventing the formation of a low-lying tower and suppressing spontaneous breaking of the SU(2) symmetry---indeed, it is not even a symmetry of the TCI Hamiltonian. This energetic isolation exposes the singlet ground states that are otherwise obscured by symmetry-breaking instabilities in short-range systems. Interestingly, such symmetric states exhibit long-range connected correlations that enable Heisenberg-limited precision in quantum sensing  \cite{block2024scalable,kaubruegger2025lieb}, making them a promising resource for quantum metrology. 

Finally, we note that many-body perturbation theory is generally singular in the thermodynamic limit: the bandwidth of each sector grows extensively with $N$, leading to spectral overlap and the eventual breakdown of the controlled expansion \cite{bravyi2011schrieffer}. Formally, maintaining the strict validity of the projected effective Hamiltonian in Eq.\,\eqref{eq:Projected_Hamiltonian} requires spectral isolation between sectors, a condition that can be enforced by rescaling the coupling as  $\lambda \!\to\! N\bar{\lambda}$ (or $J_{ij} \!\to\! \bar{J}_{ij}/N$). Importantly, however, the loss of formal control over the expansion does not necessarily signal the disappearance of the associated phase: the low-energy physics can remain adiabatically connected to that of the controlled regime even after spectral overlap develops. In the TCI model, we find that the outcome depends sensitively on the nature of the ground states of the classical Ising Hamiltonian, as discussed below.

\begin{figure}[t]
\centering
\includegraphics[width=0.48\textwidth]{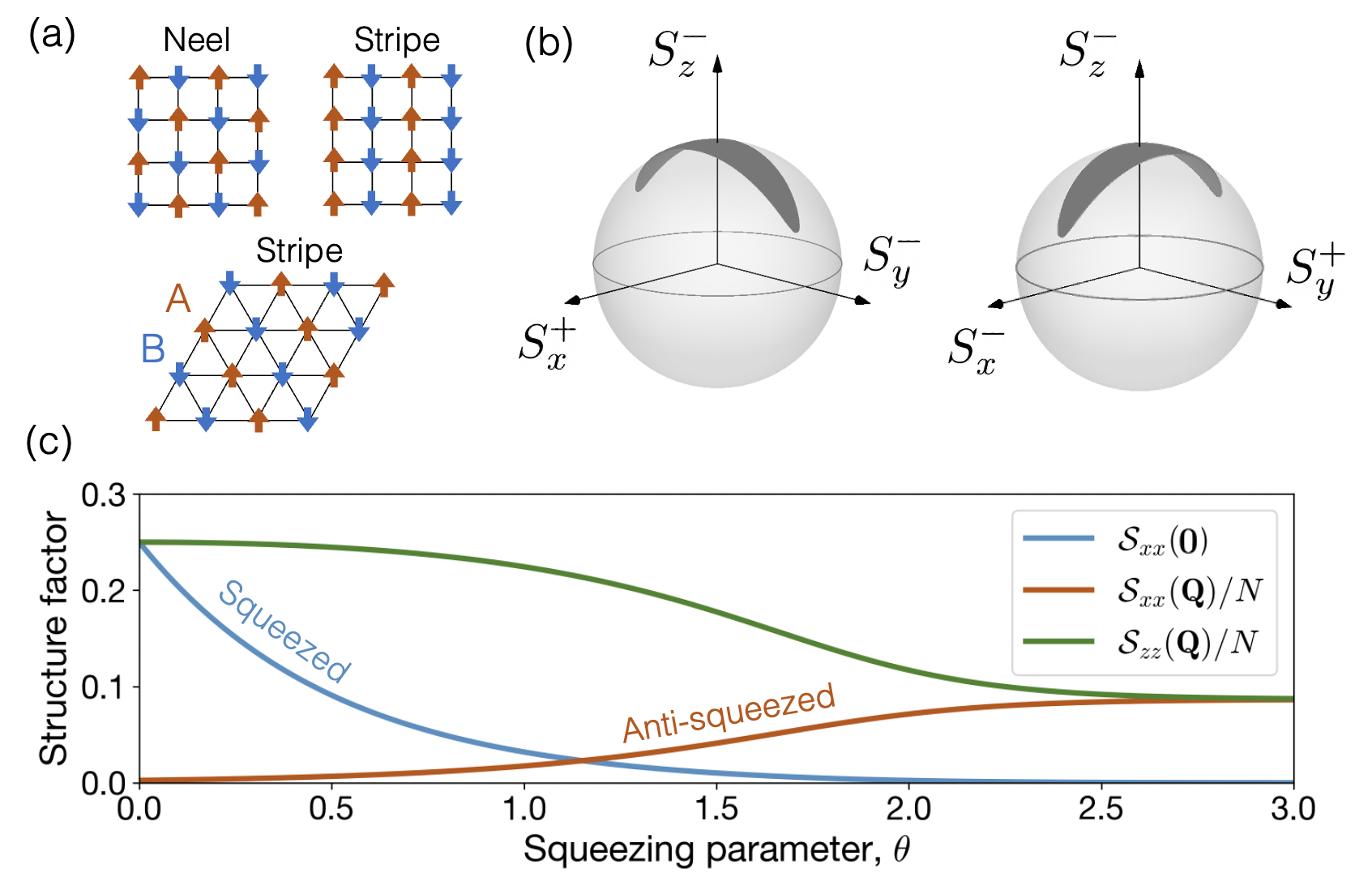}
\caption{\textbf{Squeezed AFM ansatz.} (a) Classical AFM ground states of the $J_1$-$J_2$ Ising model on the square and triangular lattices. Each state can be written as a product of two macroscopic sublattice spins ($A$ and $B$), polarized in opposite directions (b) Squeezed AFM state visualized as correlated noise distributions on a pair of generalized Bloch spheres, where the joint spin operators are $S_\alpha^\pm=S_\alpha^A \pm S_\alpha^B$. (c) Static spin structure factors as functions of the squeezing parameter. Transverse ferromagnetic correlations (blue line) are squeezed, transverse antiferromagnetic correlations (red line) are anti-squeezed, and longitudinal antiferromagnetic correlations (green line) are partially suppressed.}
\label{fig:Squeezed_AFM}
\end{figure}

\vspace{1mm}
\textbf{\textit{Squeezed AFM ansatz}}.--- Away from the strong-cavity limit, the exact Heisenberg mapping no longer applies, opening the possibility for novel quantum phases and phase transitions. To map out the rest of the ground-state phase diagrams, we first build intuition from the classical Ising limit ($\lambda=0$), focusing first on the square and triangular lattices --- the more frustrated kagome lattice will be discussed later. On the square lattice, the exact ground state of the $J_1$-$J_2$ Ising model is a classical Néel AFM ($\mathbf{Q}=\mathrm{M}$) for $J_2/J_1 < 0.5$ and a classical stripe AFM ($\mathbf{Q}=\mathrm{M}$) for $J_2/J_1 > 0.5$, while on the triangular lattice, a classical stripe AFM is realized for any $J_2>0$ (Fig.\,\ref{fig:Squeezed_AFM}\textcolor{RoyalBlue}{a}).

To construct a suitable variational ansatz in the weak-cavity limit $\lambda/J_{ij}\to 0 $, we note that these classical AFM states can be expressed as a tensor product of two macroscopic spin states $\ket{\mathrm{AFM}}= \ket{\uparrow^{\otimes N/2}}_{A}\ket{\downarrow^{\otimes N/2}}_{B}$, where the $A$ and $B$ sublattices contain spins polarized in opposite directions. Introducing a finite cavity coupling ($\lambda>0$) penalizes fluctuations in the transverse components of the collective spin $S_\alpha^A+S_\alpha^B$ ($\alpha=x,y$), where $S_\alpha^A=\sum_{i\in A }S_\alpha^i$ and $S_\alpha^B=\sum_{i\in B }S_\alpha^i$ are the sublattice spin operators. For unentangled product states like $\ket{\mathrm{AFM}}$, these fluctuations scale with system size, leading to an extensive energy penalty. To lower this penalty, the system will develop EPR-like entanglement between the two sublattice spins, such that their transverse fluctuations become anti-correlated.

%------------ FIGURE 3 -------------% 
%------------ FIGURE 3 -------------%

This motivates the following variational ansatz in the weak-cavity limit
%----------------------
\begin{equation}
    \ket{\theta} =\mathcal{A}\sum_{n=0}^{2s} (-\tanh\theta)^n \ket{s-n}_A\ket{n-s}_B \,,
    \label{eq:SqueezedAFM}
\end{equation}
%----------------------
which is the spin generalization of the bosonic two-mode squeezed vacuum state \cite{cable2010parameter,kitzinger2020two,sundar2023bosonic,mamaev2025non,kaubruegger2025lieb}. Here, $\theta$ is the squeezing parameter, $\mathcal{A}$ is a normalization factor, and $\ket{m}_{A/B}$ denote fully symmetric Dicke states on sublattices $A$ and $B$, with $s = N/4$ and $|m|\leq s$ (see Methods). These squeezed AFM states can be visualized on a pair of generalized Bloch spheres (Fig.\,\ref{fig:Squeezed_AFM}\textcolor{RoyalBlue}{b}), where the axes correspond to the joint spin operators $S_\alpha^\pm=S_\alpha^A \pm S_\alpha^B$, and the squeezed/anti-squeezed quadratures are naturally encoded in the structure factors $\mathcal{S}_{\alpha\alpha}(\mathbf{0})=\langle (S_\alpha^+)^2 \rangle_\theta/N$ and $\mathcal{S}_{\alpha\alpha}(\mathbf{Q})=\langle (S_\alpha^-)^2 \rangle_\theta/N$.

%------------ FIGURE 4 -------------% 
\begin{figure*}[t]
\centering
\includegraphics[width=\textwidth]{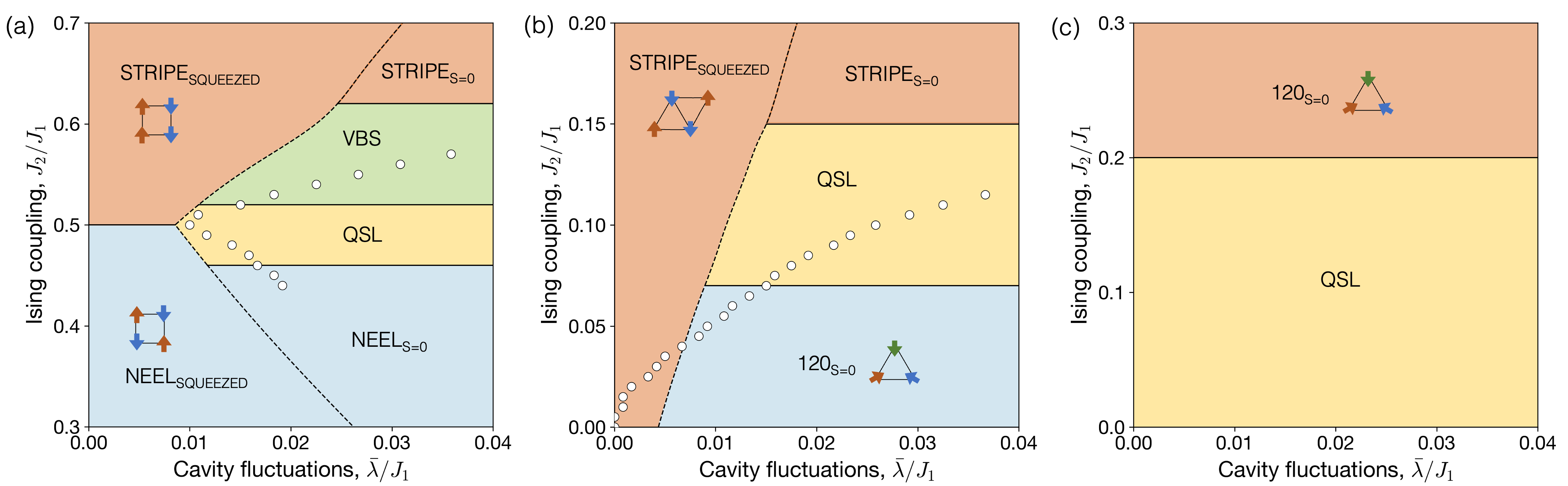}
\caption{\textbf{Ground-state phase diagrams.} Variational ground-state phase diagram of the $J_1$-$J_2$ TCI model on the (a) square and (b) triangular lattices, shown as a function of $J_2/J_1$ and $\bar{\lambda}/J_1$. In the strong-cavity limit $\bar{\lambda}/J_1\to\infty$, we expect the same singlet ground state as the corresponding Heisenberg model, and the solid lines indicate the approximate Heisenberg phase boundaries. The dashed lines indicate the critical points where the variational energies of squeezed AFM and Heisenberg singlet states crossover, $E(\theta^\star)=E_{\mathrm{Heis}}$. White dots indicate peaks in the fidelity susceptibility obtain from ED calculations on 36-site clusters (Extended Data Fig.\,\ref{fig:ED_Square},\ref{fig:ED_Triangular}).  (c) Conjectured phase diagram on the kagome lattice based on Heisenberg mapping and ED calculations which show no signature of a phase transition as $\bar{\lambda}/J_1$ is reduced (Extended Data Fig.\,\ref{fig:ED_Kagome}).}
\label{fig:Phase_Diagrams}
\end{figure*}
%------------ FIGURE 4 -------------%

The variational energy can then be expressed as
%----------------------
\begin{equation}
    E(\theta)  = \lambda N [ \mathcal{S}_{xx}(\mathbf{0}) + \mathcal{S}_{yy}(\mathbf{0})] -2J\mathcal{S}_{zz}(\mathbf{Q})\,,
    \label{eq:Variational_Energy}
\end{equation}
%----------------------
where $J=J_1-J_2$ for the square Néel AFM, $J=J_2$ for the square stripe AFM, and $J=(J_1+J_2)/2$ for the triangular stripe AFM (see Methods). As $\theta > 0$ increases, the transverse ferromagnetic components $\mathcal{S}_{xx}(\mathbf{0}) = \mathcal{S}_{yy}(\mathbf{0})$ are squeezed, while the transverse antiferromagnetic correlations $\mathcal{S}_{xx}(\mathbf{Q}) = \mathcal{S}_{yy}(\mathbf{Q})$ are anti-squeezed (Fig.\,\ref{fig:Squeezed_AFM}\textcolor{RoyalBlue}{c}), thereby lowering the cavity-induced energy penalty. However, the longitudinal antiferromagnetic correlations $\mathcal{S}_{zz}(\mathbf{Q})$ also decrease, which is penalized by the Ising interactions. 

In the large squeezing limit $\theta\to\infty$, the ansatz asymptotically approaches the singlet-projected AFM state $\ket{\theta}\to\mathbb{P}_{0,0}\ket{\mathrm{AFM}}$. This highly entangled state can be represented as an equal-weight superposition of classical AFM states oriented in all directions, or equivalently, as a fully symmetrized singlet in which each spin on sublattice $A$ forms a singlet with every spin on sublattice $B$. However, in this limit, we know that the singlet ground state of the corresponding Heisenberg model---featuring richer spatial correlations---ultimately yields a lower variational energy. 

To gain analytical insight into the transition point, we apply a Dyson-Maleev transformation to the sublattice spins, which is valid in the weak-squeezing regime (see Methods). This yields an optimal squeezing parameter that is asymptotically given by $\theta^\star \sim  \ln(\lambda N/2J)^{1/4}$ in the large-$N$ limit, and a corresponding minimum variational energy $ E(\theta^\star) \sim  - JN/2+\sqrt{2J\lambda N}$. On the other hand, the variational energy of the Heisenberg singlet ground state can be generally written as $E_{\mathrm{Heis}}=-AJN/2$, where $A<1$. Crucially, the cavity-induced energy penalty scales sub-extensively $\sim\sqrt{N}$, implying that the critical ratio $\lambda/J$ separating the two regimes increases with $N$. Therefore, in the $N\to\infty$ limit, the squeezed AFM state remains energetically favorable compared to the Heisenberg singlet ansatz for any finite $\lambda>0$. 

Under the rescaling $\lambda \to N \bar{\lambda}$, the optimal squeezing parameter is asymptotically given by $\theta^\star \sim  \ln(\eta N^2)^{1/4}$, where $\eta= \bar{\lambda}/J\beta^2$ and $\beta(\bar{\lambda}/J)$ is given in the Methods. In this regime, the longitudinal antiferromagnetic correlations are reduced from the classical value as $\mathcal{S}_{zz}(\mathbf{Q})\sim (N/4)[1-2(\sqrt{\eta}-\eta)]$, the transverse antiferromagnetic correlations grow extensively as $\mathcal{S}_{xx}(\mathbf{Q})\sim (N/4)(\sqrt{\eta}-\eta)$, while the corresponding minimum energy is
%----------------------
\begin{equation}
     E(\theta^\star)  \sim -\frac{1}{2}JN+\frac{1}{2}N(\beta+\beta^{-1})\sqrt{2J\bar{\lambda}}-\frac{1}{2}\bar{\lambda}N(1+\beta^{-2})\,.
     \label{eq:Min_Variational_Energy}
\end{equation}
%----------------------
Consequently, there exists a critical value $\bar{\lambda}/J>0$ beyond which the Heisenberg singlet ground state has a lower variational energy. This analysis suggests that, when the Ising Hamiltonian favors a classical AFM state, the appearance of the Heisenberg singlet is intrinsically tied to the validity of the projected effective Hamiltonian.

Finally, we note that the cavity-mediated interactions lift the degeneracy of the classical AFM ground states and suppress the spontaneous breaking of the discrete $Z_2$ spin-inversion symmetry. To quantify this effect, we performed a variational calculation with the symmetrized squeezed AFM states $\ket{\theta,\pm}\sim (1\pm \Pi)\ket{\theta}$, where $\Pi$ is the global spin-inversion operator, and showed that the variational gap scales extensively $\sim N$ (see Methods). This behavior aligns with our expectation in the strong-cavity limit, since these symmetrized states asymptotically approach different total-spin sectors: one becomes a singlet ($S = 0$) and the other a triplet ($S = 1$), reflecting their opposite parity $\Pi \ket{\theta,\pm}=\pm \ket{\theta,\pm}$. In contrast, the cavity does not prevent the spontaneous breaking of lattice space-group symmetries. For instance, the cavity reduces the fourfold (sixfold) degeneracy of the classical stripe AFM on the square (triangular) lattice, to a twofold (threefold) quasi-degeneracy associated with the choice of stripe orientation for a fixed parity.

\vspace{1mm}
\noindent \textbf{\textit{Ground-state phase diagrams}}.---  We can now construct a variational ground-state phase diagram for the square and triangular lattices (Fig.\,\ref{fig:Phase_Diagrams}\textcolor{RoyalBlue}{a,b}). The solid horizontal lines mark the approximate phase boundaries of the Heisenberg models \cite{jiang2012spin,hu2015competing,kolley2015phase}, while the diagonal dashed lines denote the critical points where the variational energies of the squeezed AFM ansatz and the Heisenberg singlet ansatz cross, $E(\theta^\star)=E_{\mathrm{Heis}}$. The Heisenberg energies $E_{\mathrm{Heis}}$ are taken from extrapolated ED and variational results \cite{richter2010spin,iqbal2016spin}.

On the square lattice, we predict that for $J_2/J_1\lesssim 0.45$ ($J_2/J_1\gtrsim 0.6$), the system undergoes a crossover from the squeezed Néel (stripe) AFM to the Heisenberg Néel (stripe) singlet, characterized by a smooth evolution of the structure factors with $\mathbf{Q}=\mathrm{M}$ ($\mathbf{Q}=\mathrm{X}$). Most strikingly, in the intermediate regime, $0.45 \lesssim J_2/J_1 \lesssim 0.6$, the global cavity fluctuations initially squeeze the classical AFM states, and then melt them into QSL (or VBS) at a critical coupling strength.

To numerically verify these predictions, we performed ED calculations on a 36-site cluster (see Methods). The resulting structure factors reproduce the expected ordering patterns and squeezed AFM correlations, and we find no clear signatures of additional intermediate phases (Extended Data Fig.\,\ref{fig:ED_Square}). We also computed the fidelity susceptibility $\chi_F=-2\ln\braket{\psi(\lambda)}{\psi(\lambda+\delta\lambda)}/(\delta\lambda)^2$ \cite{you2007fidelity}, which quantifies the sensitivity of the ground state $\ket{\psi(\lambda)}$ to a small change in the cavity coupling $\delta\lambda$, and identified peaks that serve as potential indicators of phase transitions (white dots in Figs.\,\ref{fig:Phase_Diagrams}\textcolor{RoyalBlue}{a,b}). As expected, these only qualitatively track the variational phase boundaries, reflecting both finite-size effects and the simplicity of the squeezed AFM ansatz that contains spatially uniform correlations.

At the maximally frustrated point $J_2/J_1 = 0.5$, the Ising model exhibits an exponentially large manifold of degenerate ground states, consisting of all classical spin configurations with zero net magnetization on each plaquette. A squeezed version of any such state yields the same variational energy, offering no indication of how quantum fluctuations lift the degeneracy. Nevertheless, our numerical results suggest that a finite cavity interaction is required to resolve this degeneracy and drive the system into the QSL phase.

On the triangular lattice for $J_2/J_1\gtrsim  0.15$, we predict a similar crossover from the squeezed stripe AFM to the Heisenberg stripe singlet, characterized by a smooth evolution of the structure factors with $\mathbf{Q}=\mathrm{M}$. In contrast, for $J_2/J_1\lesssim 0.07$ we predict a phase transition to the Heisenberg $120^\circ$ singlet, characterized by an abrupt change in the structure factors with $\mathbf{Q}=\mathrm{K}$. Most interestingly, in the intermediate regime, $0.07\lesssim  J_2/J_1\lesssim 0.15$, the global cavity interactions once again squeeze the classical AFM state and ultimately melt them into a QSL. 

The ED results for a 36-site cluster are again consistent with this general phase structure (Extended Data Fig.\,\ref{fig:ED_Triangular}). However, a qualitative difference is observed near the maximally frustrated point $J_2=0$. Our squeezed AFM ansatz does not apply here, as there is an exponentially large number of degenerate ground states in the Ising limit, comprised of all classical spin configurations satisfying the two-up-one-down (or two-down-one-up) constraint in each plaquette. However, the numerics reveal that an infinitesimal cavity perturbation resolves this degeneracy, and immediately selects a ground state which is smoothly connected to the $120^\circ$ Heisenberg singlet in the strong-cavity limit. Notably, a recent DMRG study revealed a similar behavior for the short-range $J_1$-$J_2$ XXZ model \cite{gallegos2025phase}.

On the kagome lattice, the weak-cavity limit presents a more subtle scenario, since the Ising model exhibits an exponentially large manifold of degenerate classical ground states for all $J_2 \geq 0$ \cite{colbois2022partial}. Consequently, our squeezed AFM ansatz does not apply directly at any point. Interestingly, our ED calculations on a 36-site cluster reveal no signature of a phase transition as the cavity interaction $\lambda$ is reduced from the strong-cavity limit: the structure factors show no appreciable change and the fidelity susceptibility is relatively featureless, only peaking at $\lambda=0$ (Extended Data Fig.\,\ref{fig:ED_Kagome}). 

These results indicate that an infinitesimal cavity perturbation is sufficient to lift the classical degeneracy and stabilize a QSL, which is smoothly connected to the Heisenberg QSL in the strong-cavity limit (Fig.\,\ref{fig:Phase_Diagrams}\textcolor{RoyalBlue}{c}). This scenario also mirrors that of the short-range $J_1$-$J_2$ XXZ model on a kagome lattice, where DMRG studies have shown that a robust QSL phase persists continuously from the easy-axis to the easy-plane limits \cite{he2015distinct}. Therefore, this represents a qualitatively distinct regime in which the projective action of the cavity, and thus the rescaling, is not essential for realizing a QSL.

\vspace{1mm}
\noindent \textbf{\textit{Discussion and outlook}}.--- We conclude by highlighting several compelling directions for future exploration. Despite the partial Heisenberg correspondence, the full nature of the cavity-induced QSLs uncovered here remains an open question. Indeed, our conventional understanding of QSLs is deeply rooted in short-range models, where a rigorous notion of locality---formalized through Lieb–Robinson bounds \cite{lieb1972finite}---underpins many foundational results, including the stability of topological order \cite{bravyi2010topological}. Similarly, the parton construction and projective symmetry group provide a powerful framework for classifying QSLs in terms of emergent gauge fields and fractionalized spinon excitations \cite{savary2016quantum}. Whether these frameworks can be meaningfully extended to nonlocal systems---or whether entirely new conceptual tools are required---remains a fascinating open frontier.

The cavity projection mechanism introduced here applies to arbitrary perturbations on any lattice in any dimension---including ferromagnetic couplings, longer-range interactions, and even multi-spin terms---implying a vast and largely unexplored design space for engineering correlated singlet states. For instance, adding third-neighbor Ising couplings on the kagome lattice could stabilize a chiral QSL that spontaneously breaks time-reversal symmetry, corresponding to a bosonic fractional quantum Hall state \cite{gong2015global,gong2014emergent}. More ambitiously, one might ask: given a target singlet state, can one systematically identify the perturbation required to realize it? In addition, multimode cavities \cite{marsh2024entanglement,kroeze2025directly,schine2016synthetic,clark2020observation} offer a natural way to enrich the landscape of strongly correlated states that can be realized via this cavity projection mechanism.

Finally, understanding the role of correlated dissipation in this cavity QED setting presents both a challenge and an opportunity. Interestingly, the QSL ground states and singlet excitations reside in the dark subspace of the cavity, so photon emission must be tied to finite-spin excitations such as fractional spinons. This opens promising avenues for partially suppressing the deleterious effects of dissipation, heralding errors during state preparation, and probing strongly correlated dynamics through the cavity output.

\begin{acknowledgements}

\noindent The authors gratefully acknowledge the computing time provided by the NHR Center NHR@SW at Johannes Gutenberg University Mainz. ED calculations were performed using the XDiag.jl library. The authors also acknowledge the hospitality of the Kavli Institute for Theoretical Physics (NSF PHY-2309135). C.-R.M. was supported by the Marie Skłodowska-Curie Actions (MSCA) Postdoctoral Fellowship ATOMAG (No. 101068503). M.O. was supported by a MSCA-COFUND PhD fellowship (No. 101081441). B.J. was supported by the MSCA Postdoctoral Fellowship QUINTO (No.101145886). G.C. was supported by the MSCA Postdoctoral Fellowship QUANLUX (No. 882536), the Quantum Technology Flagship project PASQuanS2.1, the WCRI Quantum Computing and Simulation Center (QCSC) of Padova University, and the Italian Ministry of University and Research MUR Departments of Excellence grant 2023-2027 Quantum Frontiers. J.M. acknowledges support from the European Union Horizon 2020 research and innovation programme under the QuantERA II project QuSiED (No 101017733); the Deutsche Forschungsgemeinschaft (No 499037529); and the Dynamics and Topology Centre, funded by the State of Rhineland Palatinate. K.S.C. acknowledges support from DND IDEAS MicroNET (MN4-056), ISED ISC Challenge (No 017137), and CFI Innovation Funds (No. 36255). D.E.C. acknowledges support from the European Union, under European Research Council grant NEWSPIN (No 101002107), EIC Pathfinder Grant PANDA (No 101115420); the Government of Spain (Project PID2024-158422NB-I00 funded by MICIU/AEI/ 10.13039/501100011033 FEDER and Severo Ochoa Grant CEX2024-001490-S [MICIU/AEI/10.13039/501100011033]); QuantERA II project QuSiED, co-funded by the European Union Horizon 2020 research and innovation programme (No 101017733) and the Government of Spain (European Union NextGenerationEU/PRTR PCI2022-132945 funded by MCIN/AEI/10.13039/501100011033); Generalitat de Catalunya (CERCA program and AGAUR Project No. 2021 SGR 01442); Fundació Cellex, and Fundació Mir-Puig.

\end{acknowledgements}

\bibliography{Main_Text}

\begin{thebibliography}{89}%
\makeatletter
\providecommand \@ifxundefined [1]{%
 \@ifx{#1\undefined}
}%
\providecommand \@ifnum [1]{%
 \ifnum #1\expandafter \@firstoftwo
 \else \expandafter \@secondoftwo
 \fi
}%
\providecommand \@ifx [1]{%
 \ifx #1\expandafter \@firstoftwo
 \else \expandafter \@secondoftwo
 \fi
}%
\providecommand \natexlab [1]{#1}%
\providecommand \enquote  [1]{``#1''}%
\providecommand \bibnamefont  [1]{#1}%
\providecommand \bibfnamefont [1]{#1}%
\providecommand \citenamefont [1]{#1}%
\providecommand \href@noop [0]{\@secondoftwo}%
\providecommand \href [0]{\begingroup \@sanitize@url \@href}%
\providecommand \@href[1]{\@@startlink{#1}\@@href}%
\providecommand \@@href[1]{\endgroup#1\@@endlink}%
\providecommand \@sanitize@url [0]{\catcode `\\12\catcode `\$12\catcode `\&12\catcode `\#12\catcode `\^12\catcode `\_12\catcode `\%12\relax}%
\providecommand \@@startlink[1]{}%
\providecommand \@@endlink[0]{}%
\providecommand \url  [0]{\begingroup\@sanitize@url \@url }%
\providecommand \@url [1]{\endgroup\@href {#1}{\urlprefix }}%
\providecommand \urlprefix  [0]{URL }%
\providecommand \Eprint [0]{\href }%
\providecommand \doibase [0]{https://doi.org/}%
\providecommand \selectlanguage [0]{\@gobble}%
\providecommand \bibinfo  [0]{\@secondoftwo}%
\providecommand \bibfield  [0]{\@secondoftwo}%
\providecommand \translation [1]{[#1]}%
\providecommand \BibitemOpen [0]{}%
\providecommand \bibitemStop [0]{}%
\providecommand \bibitemNoStop [0]{.\EOS\space}%
\providecommand \EOS [0]{\spacefactor3000\relax}%
\providecommand \BibitemShut  [1]{\csname bibitem#1\endcsname}%
\let\auto@bib@innerbib\@empty
%</preamble>
\bibitem [{\citenamefont {Leroux}\ \emph {et~al.}(2010)\citenamefont {Leroux}, \citenamefont {Schleier-Smith},\ and\ \citenamefont {Vuleti{\'c}}}]{leroux2010implementation}%
  \BibitemOpen
  \bibfield  {author} {\bibinfo {author} {\bibfnamefont {I.~D.}\ \bibnamefont {Leroux}}, \bibinfo {author} {\bibfnamefont {M.~H.}\ \bibnamefont {Schleier-Smith}},\ and\ \bibinfo {author} {\bibfnamefont {V.}~\bibnamefont {Vuleti{\'c}}},\ }\bibfield  {title} {\bibinfo {title} {{Implementation of cavity squeezing of a collective atomic spin}},\ }\href@noop {} {\bibfield  {journal} {\bibinfo  {journal} {Physical Review Letters}\ }\textbf {\bibinfo {volume} {104}},\ \bibinfo {pages} {073602} (\bibinfo {year} {2010})}\BibitemShut {NoStop}%
\bibitem [{\citenamefont {Bohnet}\ \emph {et~al.}(2014)\citenamefont {Bohnet}, \citenamefont {Cox}, \citenamefont {Norcia}, \citenamefont {Weiner}, \citenamefont {Chen},\ and\ \citenamefont {Thompson}}]{bohnet2014reduced}%
  \BibitemOpen
  \bibfield  {author} {\bibinfo {author} {\bibfnamefont {J.~G.}\ \bibnamefont {Bohnet}}, \bibinfo {author} {\bibfnamefont {K.~C.}\ \bibnamefont {Cox}}, \bibinfo {author} {\bibfnamefont {M.~A.}\ \bibnamefont {Norcia}}, \bibinfo {author} {\bibfnamefont {J.~M.}\ \bibnamefont {Weiner}}, \bibinfo {author} {\bibfnamefont {Z.}~\bibnamefont {Chen}},\ and\ \bibinfo {author} {\bibfnamefont {J.~K.}\ \bibnamefont {Thompson}},\ }\bibfield  {title} {\bibinfo {title} {{Reduced spin measurement back-action for a phase sensitivity ten times beyond the standard quantum limit}},\ }\href@noop {} {\bibfield  {journal} {\bibinfo  {journal} {Nature Photonics}\ }\textbf {\bibinfo {volume} {8}},\ \bibinfo {pages} {731} (\bibinfo {year} {2014})}\BibitemShut {NoStop}%
\bibitem [{\citenamefont {Hosten}\ \emph {et~al.}(2016)\citenamefont {Hosten}, \citenamefont {Engelsen}, \citenamefont {Krishnakumar},\ and\ \citenamefont {Kasevich}}]{hosten2016measurement}%
  \BibitemOpen
  \bibfield  {author} {\bibinfo {author} {\bibfnamefont {O.}~\bibnamefont {Hosten}}, \bibinfo {author} {\bibfnamefont {N.~J.}\ \bibnamefont {Engelsen}}, \bibinfo {author} {\bibfnamefont {R.}~\bibnamefont {Krishnakumar}},\ and\ \bibinfo {author} {\bibfnamefont {M.~A.}\ \bibnamefont {Kasevich}},\ }\bibfield  {title} {\bibinfo {title} {{Measurement noise 100 times lower than the quantum-projection limit using entangled atoms}},\ }\href@noop {} {\bibfield  {journal} {\bibinfo  {journal} {Nature}\ }\textbf {\bibinfo {volume} {529}},\ \bibinfo {pages} {505} (\bibinfo {year} {2016})}\BibitemShut {NoStop}%
\bibitem [{\citenamefont {Cox}\ \emph {et~al.}(2016)\citenamefont {Cox}, \citenamefont {Greve}, \citenamefont {Weiner},\ and\ \citenamefont {Thompson}}]{cox2016deterministic}%
  \BibitemOpen
  \bibfield  {author} {\bibinfo {author} {\bibfnamefont {K.~C.}\ \bibnamefont {Cox}}, \bibinfo {author} {\bibfnamefont {G.~P.}\ \bibnamefont {Greve}}, \bibinfo {author} {\bibfnamefont {J.~M.}\ \bibnamefont {Weiner}},\ and\ \bibinfo {author} {\bibfnamefont {J.~K.}\ \bibnamefont {Thompson}},\ }\bibfield  {title} {\bibinfo {title} {{Deterministic squeezed states with collective measurements and feedback}},\ }\href@noop {} {\bibfield  {journal} {\bibinfo  {journal} {Physical Review Letters}\ }\textbf {\bibinfo {volume} {116}},\ \bibinfo {pages} {093602} (\bibinfo {year} {2016})}\BibitemShut {NoStop}%
\bibitem [{\citenamefont {Pedrozo-Pe{\~n}afiel}\ \emph {et~al.}(2020)\citenamefont {Pedrozo-Pe{\~n}afiel}, \citenamefont {Colombo}, \citenamefont {Shu}, \citenamefont {Adiyatullin}, \citenamefont {Li}, \citenamefont {Mendez}, \citenamefont {Braverman}, \citenamefont {Kawasaki}, \citenamefont {Akamatsu}, \citenamefont {Xiao} \emph {et~al.}}]{pedrozo2020entanglement}%
  \BibitemOpen
  \bibfield  {author} {\bibinfo {author} {\bibfnamefont {E.}~\bibnamefont {Pedrozo-Pe{\~n}afiel}}, \bibinfo {author} {\bibfnamefont {S.}~\bibnamefont {Colombo}}, \bibinfo {author} {\bibfnamefont {C.}~\bibnamefont {Shu}}, \bibinfo {author} {\bibfnamefont {A.~F.}\ \bibnamefont {Adiyatullin}}, \bibinfo {author} {\bibfnamefont {Z.}~\bibnamefont {Li}}, \bibinfo {author} {\bibfnamefont {E.}~\bibnamefont {Mendez}}, \bibinfo {author} {\bibfnamefont {B.}~\bibnamefont {Braverman}}, \bibinfo {author} {\bibfnamefont {A.}~\bibnamefont {Kawasaki}}, \bibinfo {author} {\bibfnamefont {D.}~\bibnamefont {Akamatsu}}, \bibinfo {author} {\bibfnamefont {Y.}~\bibnamefont {Xiao}}, \emph {et~al.},\ }\bibfield  {title} {\bibinfo {title} {{Entanglement on an optical atomic-clock transition}},\ }\href@noop {} {\bibfield  {journal} {\bibinfo  {journal} {Nature}\ }\textbf {\bibinfo {volume} {588}},\ \bibinfo {pages} {414} (\bibinfo {year} {2020})}\BibitemShut {NoStop}%
\bibitem [{\citenamefont {Cooper}\ \emph {et~al.}(2024)\citenamefont {Cooper}, \citenamefont {Kunkel}, \citenamefont {Periwal},\ and\ \citenamefont {Schleier-Smith}}]{cooper2024graph}%
  \BibitemOpen
  \bibfield  {author} {\bibinfo {author} {\bibfnamefont {E.~S.}\ \bibnamefont {Cooper}}, \bibinfo {author} {\bibfnamefont {P.}~\bibnamefont {Kunkel}}, \bibinfo {author} {\bibfnamefont {A.}~\bibnamefont {Periwal}},\ and\ \bibinfo {author} {\bibfnamefont {M.}~\bibnamefont {Schleier-Smith}},\ }\bibfield  {title} {\bibinfo {title} {{Graph states of atomic ensembles engineered by photon-mediated entanglement}},\ }\href@noop {} {\bibfield  {journal} {\bibinfo  {journal} {Nature Physics}\ }\textbf {\bibinfo {volume} {20}},\ \bibinfo {pages} {770} (\bibinfo {year} {2024})}\BibitemShut {NoStop}%
\bibitem [{\citenamefont {Robinson}\ \emph {et~al.}(2024)\citenamefont {Robinson}, \citenamefont {Miklos}, \citenamefont {Tso}, \citenamefont {Kennedy}, \citenamefont {Bothwell}, \citenamefont {Kedar}, \citenamefont {Thompson},\ and\ \citenamefont {Ye}}]{robinson2024direct}%
  \BibitemOpen
  \bibfield  {author} {\bibinfo {author} {\bibfnamefont {J.~M.}\ \bibnamefont {Robinson}}, \bibinfo {author} {\bibfnamefont {M.}~\bibnamefont {Miklos}}, \bibinfo {author} {\bibfnamefont {Y.~M.}\ \bibnamefont {Tso}}, \bibinfo {author} {\bibfnamefont {C.~J.}\ \bibnamefont {Kennedy}}, \bibinfo {author} {\bibfnamefont {T.}~\bibnamefont {Bothwell}}, \bibinfo {author} {\bibfnamefont {D.}~\bibnamefont {Kedar}}, \bibinfo {author} {\bibfnamefont {J.~K.}\ \bibnamefont {Thompson}},\ and\ \bibinfo {author} {\bibfnamefont {J.}~\bibnamefont {Ye}},\ }\bibfield  {title} {\bibinfo {title} {{Direct comparison of two spin-squeezed optical clock ensembles at the $10^{-17}$ level}},\ }\href@noop {} {\bibfield  {journal} {\bibinfo  {journal} {Nature Physics}\ }\textbf {\bibinfo {volume} {20}},\ \bibinfo {pages} {208} (\bibinfo {year} {2024})}\BibitemShut {NoStop}%
\bibitem [{\citenamefont {Baumann}\ \emph {et~al.}(2010)\citenamefont {Baumann}, \citenamefont {Guerlin}, \citenamefont {Brennecke},\ and\ \citenamefont {Esslinger}}]{baumann2010dicke}%
  \BibitemOpen
  \bibfield  {author} {\bibinfo {author} {\bibfnamefont {K.}~\bibnamefont {Baumann}}, \bibinfo {author} {\bibfnamefont {C.}~\bibnamefont {Guerlin}}, \bibinfo {author} {\bibfnamefont {F.}~\bibnamefont {Brennecke}},\ and\ \bibinfo {author} {\bibfnamefont {T.}~\bibnamefont {Esslinger}},\ }\bibfield  {title} {\bibinfo {title} {{Dicke quantum phase transition with a superfluid gas in an optical cavity}},\ }\href@noop {} {\bibfield  {journal} {\bibinfo  {journal} {Nature}\ }\textbf {\bibinfo {volume} {464}},\ \bibinfo {pages} {1301} (\bibinfo {year} {2010})}\BibitemShut {NoStop}%
\bibitem [{\citenamefont {Bohnet}\ \emph {et~al.}(2012)\citenamefont {Bohnet}, \citenamefont {Chen}, \citenamefont {Weiner}, \citenamefont {Meiser}, \citenamefont {Holland},\ and\ \citenamefont {Thompson}}]{bohnet2012steady}%
  \BibitemOpen
  \bibfield  {author} {\bibinfo {author} {\bibfnamefont {J.~G.}\ \bibnamefont {Bohnet}}, \bibinfo {author} {\bibfnamefont {Z.}~\bibnamefont {Chen}}, \bibinfo {author} {\bibfnamefont {J.~M.}\ \bibnamefont {Weiner}}, \bibinfo {author} {\bibfnamefont {D.}~\bibnamefont {Meiser}}, \bibinfo {author} {\bibfnamefont {M.~J.}\ \bibnamefont {Holland}},\ and\ \bibinfo {author} {\bibfnamefont {J.~K.}\ \bibnamefont {Thompson}},\ }\bibfield  {title} {\bibinfo {title} {{A steady-state superradiant laser with less than one intracavity photon}},\ }\href@noop {} {\bibfield  {journal} {\bibinfo  {journal} {Nature}\ }\textbf {\bibinfo {volume} {484}},\ \bibinfo {pages} {78} (\bibinfo {year} {2012})}\BibitemShut {NoStop}%
\bibitem [{\citenamefont {Norcia}\ \emph {et~al.}(2016)\citenamefont {Norcia}, \citenamefont {Winchester}, \citenamefont {Cline},\ and\ \citenamefont {Thompson}}]{norcia2016superradiance}%
  \BibitemOpen
  \bibfield  {author} {\bibinfo {author} {\bibfnamefont {M.~A.}\ \bibnamefont {Norcia}}, \bibinfo {author} {\bibfnamefont {M.~N.}\ \bibnamefont {Winchester}}, \bibinfo {author} {\bibfnamefont {J.~R.}\ \bibnamefont {Cline}},\ and\ \bibinfo {author} {\bibfnamefont {J.~K.}\ \bibnamefont {Thompson}},\ }\bibfield  {title} {\bibinfo {title} {{Superradiance on the millihertz linewidth strontium clock transition}},\ }\href@noop {} {\bibfield  {journal} {\bibinfo  {journal} {Science Advances}\ }\textbf {\bibinfo {volume} {2}},\ \bibinfo {pages} {e1601231} (\bibinfo {year} {2016})}\BibitemShut {NoStop}%
\bibitem [{\citenamefont {Song}\ \emph {et~al.}(2025)\citenamefont {Song}, \citenamefont {Barberena}, \citenamefont {Young}, \citenamefont {Chaparro}, \citenamefont {Chu}, \citenamefont {Agarwal}, \citenamefont {Niu}, \citenamefont {Young}, \citenamefont {Rey},\ and\ \citenamefont {Thompson}}]{song2025dissipation}%
  \BibitemOpen
  \bibfield  {author} {\bibinfo {author} {\bibfnamefont {E.~Y.}\ \bibnamefont {Song}}, \bibinfo {author} {\bibfnamefont {D.}~\bibnamefont {Barberena}}, \bibinfo {author} {\bibfnamefont {D.~J.}\ \bibnamefont {Young}}, \bibinfo {author} {\bibfnamefont {E.}~\bibnamefont {Chaparro}}, \bibinfo {author} {\bibfnamefont {A.}~\bibnamefont {Chu}}, \bibinfo {author} {\bibfnamefont {S.}~\bibnamefont {Agarwal}}, \bibinfo {author} {\bibfnamefont {Z.}~\bibnamefont {Niu}}, \bibinfo {author} {\bibfnamefont {J.~T.}\ \bibnamefont {Young}}, \bibinfo {author} {\bibfnamefont {A.~M.}\ \bibnamefont {Rey}},\ and\ \bibinfo {author} {\bibfnamefont {J.~K.}\ \bibnamefont {Thompson}},\ }\bibfield  {title} {\bibinfo {title} {{A dissipation-induced superradiant transition in a strontium cavity-QED system}},\ }\href@noop {} {\bibfield  {journal} {\bibinfo  {journal} {Science Advances}\ }\textbf {\bibinfo {volume} {11}},\ \bibinfo {pages} {eadu5799} (\bibinfo {year} {2025})}\BibitemShut {NoStop}%
\bibitem [{\citenamefont {Ke{\ss}ler}\ \emph {et~al.}(2021)\citenamefont {Ke{\ss}ler}, \citenamefont {Kongkhambut}, \citenamefont {Georges}, \citenamefont {Mathey}, \citenamefont {Cosme},\ and\ \citenamefont {Hemmerich}}]{kessler2021observation}%
  \BibitemOpen
  \bibfield  {author} {\bibinfo {author} {\bibfnamefont {H.}~\bibnamefont {Ke{\ss}ler}}, \bibinfo {author} {\bibfnamefont {P.}~\bibnamefont {Kongkhambut}}, \bibinfo {author} {\bibfnamefont {C.}~\bibnamefont {Georges}}, \bibinfo {author} {\bibfnamefont {L.}~\bibnamefont {Mathey}}, \bibinfo {author} {\bibfnamefont {J.~G.}\ \bibnamefont {Cosme}},\ and\ \bibinfo {author} {\bibfnamefont {A.}~\bibnamefont {Hemmerich}},\ }\bibfield  {title} {\bibinfo {title} {{Observation of a dissipative time crystal}},\ }\href@noop {} {\bibfield  {journal} {\bibinfo  {journal} {Physical Review Letters}\ }\textbf {\bibinfo {volume} {127}},\ \bibinfo {pages} {043602} (\bibinfo {year} {2021})}\BibitemShut {NoStop}%
\bibitem [{\citenamefont {Kongkhambut}\ \emph {et~al.}(2022)\citenamefont {Kongkhambut}, \citenamefont {Skulte}, \citenamefont {Mathey}, \citenamefont {Cosme}, \citenamefont {Hemmerich},\ and\ \citenamefont {Ke{\ss}ler}}]{kongkhambut2022observation}%
  \BibitemOpen
  \bibfield  {author} {\bibinfo {author} {\bibfnamefont {P.}~\bibnamefont {Kongkhambut}}, \bibinfo {author} {\bibfnamefont {J.}~\bibnamefont {Skulte}}, \bibinfo {author} {\bibfnamefont {L.}~\bibnamefont {Mathey}}, \bibinfo {author} {\bibfnamefont {J.~G.}\ \bibnamefont {Cosme}}, \bibinfo {author} {\bibfnamefont {A.}~\bibnamefont {Hemmerich}},\ and\ \bibinfo {author} {\bibfnamefont {H.}~\bibnamefont {Ke{\ss}ler}},\ }\bibfield  {title} {\bibinfo {title} {{Observation of a continuous time crystal}},\ }\href@noop {} {\bibfield  {journal} {\bibinfo  {journal} {Science}\ }\textbf {\bibinfo {volume} {377}},\ \bibinfo {pages} {670} (\bibinfo {year} {2022})}\BibitemShut {NoStop}%
\bibitem [{\citenamefont {Muniz}\ \emph {et~al.}(2020)\citenamefont {Muniz}, \citenamefont {Barberena}, \citenamefont {Lewis-Swan}, \citenamefont {Young}, \citenamefont {Cline}, \citenamefont {Rey},\ and\ \citenamefont {Thompson}}]{muniz2020exploring}%
  \BibitemOpen
  \bibfield  {author} {\bibinfo {author} {\bibfnamefont {J.~A.}\ \bibnamefont {Muniz}}, \bibinfo {author} {\bibfnamefont {D.}~\bibnamefont {Barberena}}, \bibinfo {author} {\bibfnamefont {R.~J.}\ \bibnamefont {Lewis-Swan}}, \bibinfo {author} {\bibfnamefont {D.~J.}\ \bibnamefont {Young}}, \bibinfo {author} {\bibfnamefont {J.~R.}\ \bibnamefont {Cline}}, \bibinfo {author} {\bibfnamefont {A.~M.}\ \bibnamefont {Rey}},\ and\ \bibinfo {author} {\bibfnamefont {J.~K.}\ \bibnamefont {Thompson}},\ }\bibfield  {title} {\bibinfo {title} {{Exploring dynamical phase transitions with cold atoms in an optical cavity}},\ }\href@noop {} {\bibfield  {journal} {\bibinfo  {journal} {Nature}\ }\textbf {\bibinfo {volume} {580}},\ \bibinfo {pages} {602} (\bibinfo {year} {2020})}\BibitemShut {NoStop}%
\bibitem [{\citenamefont {Young}\ \emph {et~al.}(2024)\citenamefont {Young}, \citenamefont {Chu}, \citenamefont {Song}, \citenamefont {Barberena}, \citenamefont {Wellnitz}, \citenamefont {Niu}, \citenamefont {Sch{\"a}fer}, \citenamefont {Lewis-Swan}, \citenamefont {Rey},\ and\ \citenamefont {Thompson}}]{young2024observing}%
  \BibitemOpen
  \bibfield  {author} {\bibinfo {author} {\bibfnamefont {D.~J.}\ \bibnamefont {Young}}, \bibinfo {author} {\bibfnamefont {A.}~\bibnamefont {Chu}}, \bibinfo {author} {\bibfnamefont {E.~Y.}\ \bibnamefont {Song}}, \bibinfo {author} {\bibfnamefont {D.}~\bibnamefont {Barberena}}, \bibinfo {author} {\bibfnamefont {D.}~\bibnamefont {Wellnitz}}, \bibinfo {author} {\bibfnamefont {Z.}~\bibnamefont {Niu}}, \bibinfo {author} {\bibfnamefont {V.~M.}\ \bibnamefont {Sch{\"a}fer}}, \bibinfo {author} {\bibfnamefont {R.~J.}\ \bibnamefont {Lewis-Swan}}, \bibinfo {author} {\bibfnamefont {A.~M.}\ \bibnamefont {Rey}},\ and\ \bibinfo {author} {\bibfnamefont {J.~K.}\ \bibnamefont {Thompson}},\ }\bibfield  {title} {\bibinfo {title} {{Observing dynamical phases of BCS superconductors in a cavity QED simulator}},\ }\href@noop {} {\bibfield  {journal} {\bibinfo  {journal} {Nature}\ }\textbf {\bibinfo {volume} {625}},\ \bibinfo {pages} {679} (\bibinfo {year} {2024})}\BibitemShut {NoStop}%
\bibitem [{\citenamefont {Semeghini}\ \emph {et~al.}(2021)\citenamefont {Semeghini}, \citenamefont {Levine}, \citenamefont {Keesling}, \citenamefont {Ebadi}, \citenamefont {Wang}, \citenamefont {Bluvstein}, \citenamefont {Verresen}, \citenamefont {Pichler}, \citenamefont {Kalinowski}, \citenamefont {Samajdar} \emph {et~al.}}]{semeghini2021probing}%
  \BibitemOpen
  \bibfield  {author} {\bibinfo {author} {\bibfnamefont {G.}~\bibnamefont {Semeghini}}, \bibinfo {author} {\bibfnamefont {H.}~\bibnamefont {Levine}}, \bibinfo {author} {\bibfnamefont {A.}~\bibnamefont {Keesling}}, \bibinfo {author} {\bibfnamefont {S.}~\bibnamefont {Ebadi}}, \bibinfo {author} {\bibfnamefont {T.~T.}\ \bibnamefont {Wang}}, \bibinfo {author} {\bibfnamefont {D.}~\bibnamefont {Bluvstein}}, \bibinfo {author} {\bibfnamefont {R.}~\bibnamefont {Verresen}}, \bibinfo {author} {\bibfnamefont {H.}~\bibnamefont {Pichler}}, \bibinfo {author} {\bibfnamefont {M.}~\bibnamefont {Kalinowski}}, \bibinfo {author} {\bibfnamefont {R.}~\bibnamefont {Samajdar}}, \emph {et~al.},\ }\bibfield  {title} {\bibinfo {title} {{Probing topological spin liquids on a programmable quantum simulator}},\ }\href@noop {} {\bibfield  {journal} {\bibinfo  {journal} {Science}\ }\textbf {\bibinfo {volume} {374}},\ \bibinfo {pages} {1242} (\bibinfo {year} {2021})}\BibitemShut {NoStop}%
\bibitem [{\citenamefont {Satzinger}\ \emph {et~al.}(2021)\citenamefont {Satzinger}, \citenamefont {Liu}, \citenamefont {Smith}, \citenamefont {Knapp}, \citenamefont {Newman}, \citenamefont {Jones}, \citenamefont {Chen}, \citenamefont {Quintana}, \citenamefont {Mi}, \citenamefont {Dunsworth} \emph {et~al.}}]{satzinger2021realizing}%
  \BibitemOpen
  \bibfield  {author} {\bibinfo {author} {\bibfnamefont {K.}~\bibnamefont {Satzinger}}, \bibinfo {author} {\bibfnamefont {Y.-J.}\ \bibnamefont {Liu}}, \bibinfo {author} {\bibfnamefont {A.}~\bibnamefont {Smith}}, \bibinfo {author} {\bibfnamefont {C.}~\bibnamefont {Knapp}}, \bibinfo {author} {\bibfnamefont {M.}~\bibnamefont {Newman}}, \bibinfo {author} {\bibfnamefont {C.}~\bibnamefont {Jones}}, \bibinfo {author} {\bibfnamefont {Z.}~\bibnamefont {Chen}}, \bibinfo {author} {\bibfnamefont {C.}~\bibnamefont {Quintana}}, \bibinfo {author} {\bibfnamefont {X.}~\bibnamefont {Mi}}, \bibinfo {author} {\bibfnamefont {A.}~\bibnamefont {Dunsworth}}, \emph {et~al.},\ }\bibfield  {title} {\bibinfo {title} {{Realizing topologically ordered states on a quantum processor}},\ }\href@noop {} {\bibfield  {journal} {\bibinfo  {journal} {Science}\ }\textbf {\bibinfo {volume} {374}},\ \bibinfo {pages} {1237} (\bibinfo {year} {2021})}\BibitemShut {NoStop}%
\bibitem [{\citenamefont {Iqbal}\ \emph {et~al.}(2024)\citenamefont {Iqbal}, \citenamefont {Tantivasadakarn}, \citenamefont {Verresen}, \citenamefont {Campbell}, \citenamefont {Dreiling}, \citenamefont {Figgatt}, \citenamefont {Gaebler}, \citenamefont {Johansen}, \citenamefont {Mills}, \citenamefont {Moses} \emph {et~al.}}]{iqbal2024non}%
  \BibitemOpen
  \bibfield  {author} {\bibinfo {author} {\bibfnamefont {M.}~\bibnamefont {Iqbal}}, \bibinfo {author} {\bibfnamefont {N.}~\bibnamefont {Tantivasadakarn}}, \bibinfo {author} {\bibfnamefont {R.}~\bibnamefont {Verresen}}, \bibinfo {author} {\bibfnamefont {S.~L.}\ \bibnamefont {Campbell}}, \bibinfo {author} {\bibfnamefont {J.~M.}\ \bibnamefont {Dreiling}}, \bibinfo {author} {\bibfnamefont {C.}~\bibnamefont {Figgatt}}, \bibinfo {author} {\bibfnamefont {J.~P.}\ \bibnamefont {Gaebler}}, \bibinfo {author} {\bibfnamefont {J.}~\bibnamefont {Johansen}}, \bibinfo {author} {\bibfnamefont {M.}~\bibnamefont {Mills}}, \bibinfo {author} {\bibfnamefont {S.~A.}\ \bibnamefont {Moses}}, \emph {et~al.},\ }\bibfield  {title} {\bibinfo {title} {{Non-abelian topological order and anyons on a trapped-ion processor}},\ }\href@noop {} {\bibfield  {journal} {\bibinfo  {journal} {Nature}\ }\textbf {\bibinfo {volume} {626}},\ \bibinfo {pages} {505} (\bibinfo {year} {2024})}\BibitemShut {NoStop}%
\bibitem [{\citenamefont {Xu}\ \emph {et~al.}(2024)\citenamefont {Xu}, \citenamefont {Sun}, \citenamefont {Wang}, \citenamefont {Li}, \citenamefont {Zhu}, \citenamefont {Dong}, \citenamefont {Deng}, \citenamefont {Zhang}, \citenamefont {Chen}, \citenamefont {Wu} \emph {et~al.}}]{xu2024non}%
  \BibitemOpen
  \bibfield  {author} {\bibinfo {author} {\bibfnamefont {S.}~\bibnamefont {Xu}}, \bibinfo {author} {\bibfnamefont {Z.-Z.}\ \bibnamefont {Sun}}, \bibinfo {author} {\bibfnamefont {K.}~\bibnamefont {Wang}}, \bibinfo {author} {\bibfnamefont {H.}~\bibnamefont {Li}}, \bibinfo {author} {\bibfnamefont {Z.}~\bibnamefont {Zhu}}, \bibinfo {author} {\bibfnamefont {H.}~\bibnamefont {Dong}}, \bibinfo {author} {\bibfnamefont {J.}~\bibnamefont {Deng}}, \bibinfo {author} {\bibfnamefont {X.}~\bibnamefont {Zhang}}, \bibinfo {author} {\bibfnamefont {J.}~\bibnamefont {Chen}}, \bibinfo {author} {\bibfnamefont {Y.}~\bibnamefont {Wu}}, \emph {et~al.},\ }\bibfield  {title} {\bibinfo {title} {{Non-abelian braiding of Fibonacci anyons with a superconducting processor}},\ }\href@noop {} {\bibfield  {journal} {\bibinfo  {journal} {Nature Physics}\ }\textbf {\bibinfo {volume} {20}},\ \bibinfo {pages} {1469} (\bibinfo {year} {2024})}\BibitemShut {NoStop}%
\bibitem [{\citenamefont {Evered}\ \emph {et~al.}(2025)\citenamefont {Evered}, \citenamefont {Kalinowski}, \citenamefont {Geim}, \citenamefont {Manovitz}, \citenamefont {Bluvstein}, \citenamefont {Li}, \citenamefont {Maskara}, \citenamefont {Zhou}, \citenamefont {Ebadi}, \citenamefont {Xu} \emph {et~al.}}]{evered2025probing}%
  \BibitemOpen
  \bibfield  {author} {\bibinfo {author} {\bibfnamefont {S.~J.}\ \bibnamefont {Evered}}, \bibinfo {author} {\bibfnamefont {M.}~\bibnamefont {Kalinowski}}, \bibinfo {author} {\bibfnamefont {A.~A.}\ \bibnamefont {Geim}}, \bibinfo {author} {\bibfnamefont {T.}~\bibnamefont {Manovitz}}, \bibinfo {author} {\bibfnamefont {D.}~\bibnamefont {Bluvstein}}, \bibinfo {author} {\bibfnamefont {S.~H.}\ \bibnamefont {Li}}, \bibinfo {author} {\bibfnamefont {N.}~\bibnamefont {Maskara}}, \bibinfo {author} {\bibfnamefont {H.}~\bibnamefont {Zhou}}, \bibinfo {author} {\bibfnamefont {S.}~\bibnamefont {Ebadi}}, \bibinfo {author} {\bibfnamefont {M.}~\bibnamefont {Xu}}, \emph {et~al.},\ }\bibfield  {title} {\bibinfo {title} {{Probing the Kitaev honeycomb model on a neutral-atom quantum computer}},\ }\href@noop {} {\bibfield  {journal} {\bibinfo  {journal} {Nature}\ }\textbf {\bibinfo {volume} {645}},\ \bibinfo {pages} {341} (\bibinfo {year} {2025})}\BibitemShut {NoStop}%
\bibitem [{\citenamefont {Will}\ \emph {et~al.}(2025)\citenamefont {Will}, \citenamefont {Cochran}, \citenamefont {Rosenberg}, \citenamefont {Jobst}, \citenamefont {Eassa}, \citenamefont {Roushan}, \citenamefont {Knap}, \citenamefont {Gammon-Smith},\ and\ \citenamefont {Pollmann}}]{will2025probing}%
  \BibitemOpen
  \bibfield  {author} {\bibinfo {author} {\bibfnamefont {M.}~\bibnamefont {Will}}, \bibinfo {author} {\bibfnamefont {T.}~\bibnamefont {Cochran}}, \bibinfo {author} {\bibfnamefont {E.}~\bibnamefont {Rosenberg}}, \bibinfo {author} {\bibfnamefont {B.}~\bibnamefont {Jobst}}, \bibinfo {author} {\bibfnamefont {N.~M.}\ \bibnamefont {Eassa}}, \bibinfo {author} {\bibfnamefont {P.}~\bibnamefont {Roushan}}, \bibinfo {author} {\bibfnamefont {M.}~\bibnamefont {Knap}}, \bibinfo {author} {\bibfnamefont {A.}~\bibnamefont {Gammon-Smith}},\ and\ \bibinfo {author} {\bibfnamefont {F.}~\bibnamefont {Pollmann}},\ }\bibfield  {title} {\bibinfo {title} {{Probing non-equilibrium topological order on a quantum processor}},\ }\href@noop {} {\bibfield  {journal} {\bibinfo  {journal} {Nature}\ }\textbf {\bibinfo {volume} {645}},\ \bibinfo {pages} {348} (\bibinfo {year} {2025})}\BibitemShut {NoStop}%
\bibitem [{\citenamefont {Periwal}\ \emph {et~al.}(2021)\citenamefont {Periwal}, \citenamefont {Cooper}, \citenamefont {Kunkel}, \citenamefont {Wienand}, \citenamefont {Davis},\ and\ \citenamefont {Schleier-Smith}}]{periwal2021programmable}%
  \BibitemOpen
  \bibfield  {author} {\bibinfo {author} {\bibfnamefont {A.}~\bibnamefont {Periwal}}, \bibinfo {author} {\bibfnamefont {E.~S.}\ \bibnamefont {Cooper}}, \bibinfo {author} {\bibfnamefont {P.}~\bibnamefont {Kunkel}}, \bibinfo {author} {\bibfnamefont {J.~F.}\ \bibnamefont {Wienand}}, \bibinfo {author} {\bibfnamefont {E.~J.}\ \bibnamefont {Davis}},\ and\ \bibinfo {author} {\bibfnamefont {M.}~\bibnamefont {Schleier-Smith}},\ }\bibfield  {title} {\bibinfo {title} {{Programmable interactions and emergent geometry in an array of atom clouds}},\ }\href@noop {} {\bibfield  {journal} {\bibinfo  {journal} {Nature}\ }\textbf {\bibinfo {volume} {600}},\ \bibinfo {pages} {630} (\bibinfo {year} {2021})}\BibitemShut {NoStop}%
\bibitem [{\citenamefont {Marsh}\ \emph {et~al.}(2024)\citenamefont {Marsh}, \citenamefont {Kroeze}, \citenamefont {Ganguli}, \citenamefont {Gopalakrishnan}, \citenamefont {Keeling},\ and\ \citenamefont {Lev}}]{marsh2024entanglement}%
  \BibitemOpen
  \bibfield  {author} {\bibinfo {author} {\bibfnamefont {B.~P.}\ \bibnamefont {Marsh}}, \bibinfo {author} {\bibfnamefont {R.~M.}\ \bibnamefont {Kroeze}}, \bibinfo {author} {\bibfnamefont {S.}~\bibnamefont {Ganguli}}, \bibinfo {author} {\bibfnamefont {S.}~\bibnamefont {Gopalakrishnan}}, \bibinfo {author} {\bibfnamefont {J.}~\bibnamefont {Keeling}},\ and\ \bibinfo {author} {\bibfnamefont {B.~L.}\ \bibnamefont {Lev}},\ }\bibfield  {title} {\bibinfo {title} {{Entanglement and replica symmetry breaking in a driven-dissipative quantum spin glass}},\ }\href@noop {} {\bibfield  {journal} {\bibinfo  {journal} {Physical Review X}\ }\textbf {\bibinfo {volume} {14}},\ \bibinfo {pages} {011026} (\bibinfo {year} {2024})}\BibitemShut {NoStop}%
\bibitem [{\citenamefont {Kroeze}\ \emph {et~al.}(2025)\citenamefont {Kroeze}, \citenamefont {Marsh}, \citenamefont {Atri~Schuller}, \citenamefont {Hunt}, \citenamefont {Bourzutschky}, \citenamefont {Winer}, \citenamefont {Gopalakrishnan}, \citenamefont {Keeling},\ and\ \citenamefont {Lev}}]{kroeze2025directly}%
  \BibitemOpen
  \bibfield  {author} {\bibinfo {author} {\bibfnamefont {R.~M.}\ \bibnamefont {Kroeze}}, \bibinfo {author} {\bibfnamefont {B.~P.}\ \bibnamefont {Marsh}}, \bibinfo {author} {\bibfnamefont {D.}~\bibnamefont {Atri~Schuller}}, \bibinfo {author} {\bibfnamefont {H.~S.}\ \bibnamefont {Hunt}}, \bibinfo {author} {\bibfnamefont {A.~N.}\ \bibnamefont {Bourzutschky}}, \bibinfo {author} {\bibfnamefont {M.}~\bibnamefont {Winer}}, \bibinfo {author} {\bibfnamefont {S.}~\bibnamefont {Gopalakrishnan}}, \bibinfo {author} {\bibfnamefont {J.}~\bibnamefont {Keeling}},\ and\ \bibinfo {author} {\bibfnamefont {B.~L.}\ \bibnamefont {Lev}},\ }\bibfield  {title} {\bibinfo {title} {{Directly observing replica symmetry breaking in a vector quantum-optical spin glass}},\ }\href@noop {} {\bibfield  {journal} {\bibinfo  {journal} {Science}\ }\textbf {\bibinfo {volume} {389}},\ \bibinfo {pages} {1122} (\bibinfo {year} {2025})}\BibitemShut {NoStop}%
\bibitem [{\citenamefont {Schine}\ \emph {et~al.}(2016)\citenamefont {Schine}, \citenamefont {Ryou}, \citenamefont {Gromov}, \citenamefont {Sommer},\ and\ \citenamefont {Simon}}]{schine2016synthetic}%
  \BibitemOpen
  \bibfield  {author} {\bibinfo {author} {\bibfnamefont {N.}~\bibnamefont {Schine}}, \bibinfo {author} {\bibfnamefont {A.}~\bibnamefont {Ryou}}, \bibinfo {author} {\bibfnamefont {A.}~\bibnamefont {Gromov}}, \bibinfo {author} {\bibfnamefont {A.}~\bibnamefont {Sommer}},\ and\ \bibinfo {author} {\bibfnamefont {J.}~\bibnamefont {Simon}},\ }\bibfield  {title} {\bibinfo {title} {{Synthetic Landau levels for photons}},\ }\href@noop {} {\bibfield  {journal} {\bibinfo  {journal} {Nature}\ }\textbf {\bibinfo {volume} {534}},\ \bibinfo {pages} {671} (\bibinfo {year} {2016})}\BibitemShut {NoStop}%
\bibitem [{\citenamefont {Clark}\ \emph {et~al.}(2020)\citenamefont {Clark}, \citenamefont {Schine}, \citenamefont {Baum}, \citenamefont {Jia},\ and\ \citenamefont {Simon}}]{clark2020observation}%
  \BibitemOpen
  \bibfield  {author} {\bibinfo {author} {\bibfnamefont {L.~W.}\ \bibnamefont {Clark}}, \bibinfo {author} {\bibfnamefont {N.}~\bibnamefont {Schine}}, \bibinfo {author} {\bibfnamefont {C.}~\bibnamefont {Baum}}, \bibinfo {author} {\bibfnamefont {N.}~\bibnamefont {Jia}},\ and\ \bibinfo {author} {\bibfnamefont {J.}~\bibnamefont {Simon}},\ }\bibfield  {title} {\bibinfo {title} {{Observation of Laughlin states made of light}},\ }\href@noop {} {\bibfield  {journal} {\bibinfo  {journal} {Nature}\ }\textbf {\bibinfo {volume} {582}},\ \bibinfo {pages} {41} (\bibinfo {year} {2020})}\BibitemShut {NoStop}%
\bibitem [{\citenamefont {Kong}\ \emph {et~al.}(2021)\citenamefont {Kong}, \citenamefont {Taylor}, \citenamefont {Dong},\ and\ \citenamefont {Choi}}]{kong2021melting}%
  \BibitemOpen
  \bibfield  {author} {\bibinfo {author} {\bibfnamefont {H.}~\bibnamefont {Kong}}, \bibinfo {author} {\bibfnamefont {J.}~\bibnamefont {Taylor}}, \bibinfo {author} {\bibfnamefont {Y.}~\bibnamefont {Dong}},\ and\ \bibinfo {author} {\bibfnamefont {K.}~\bibnamefont {Choi}},\ }\bibfield  {title} {\bibinfo {title} {{Melting a Rydberg ice to a topological spin liquid with cavity vacuum fluctuation}},\ }\href@noop {} {\bibfield  {journal} {\bibinfo  {journal} {arXiv preprint arXiv:2109.03741}\ } (\bibinfo {year} {2021})}\BibitemShut {NoStop}%
\bibitem [{\citenamefont {Deist}\ \emph {et~al.}(2022)\citenamefont {Deist}, \citenamefont {Lu}, \citenamefont {Ho}, \citenamefont {Pasha}, \citenamefont {Zeiher}, \citenamefont {Yan},\ and\ \citenamefont {Stamper-Kurn}}]{deist2022mid}%
  \BibitemOpen
  \bibfield  {author} {\bibinfo {author} {\bibfnamefont {E.}~\bibnamefont {Deist}}, \bibinfo {author} {\bibfnamefont {Y.-H.}\ \bibnamefont {Lu}}, \bibinfo {author} {\bibfnamefont {J.}~\bibnamefont {Ho}}, \bibinfo {author} {\bibfnamefont {M.~K.}\ \bibnamefont {Pasha}}, \bibinfo {author} {\bibfnamefont {J.}~\bibnamefont {Zeiher}}, \bibinfo {author} {\bibfnamefont {Z.}~\bibnamefont {Yan}},\ and\ \bibinfo {author} {\bibfnamefont {D.~M.}\ \bibnamefont {Stamper-Kurn}},\ }\bibfield  {title} {\bibinfo {title} {{Mid-circuit cavity measurement in a neutral atom array}},\ }\href@noop {} {\bibfield  {journal} {\bibinfo  {journal} {Physical Review Letters}\ }\textbf {\bibinfo {volume} {129}},\ \bibinfo {pages} {203602} (\bibinfo {year} {2022})}\BibitemShut {NoStop}%
\bibitem [{\citenamefont {Liu}\ \emph {et~al.}(2023)\citenamefont {Liu}, \citenamefont {Wang}, \citenamefont {Yang}, \citenamefont {Wang}, \citenamefont {Fan}, \citenamefont {Guan}, \citenamefont {Li}, \citenamefont {Zhang},\ and\ \citenamefont {Zhang}}]{liu2023realization}%
  \BibitemOpen
  \bibfield  {author} {\bibinfo {author} {\bibfnamefont {Y.}~\bibnamefont {Liu}}, \bibinfo {author} {\bibfnamefont {Z.}~\bibnamefont {Wang}}, \bibinfo {author} {\bibfnamefont {P.}~\bibnamefont {Yang}}, \bibinfo {author} {\bibfnamefont {Q.}~\bibnamefont {Wang}}, \bibinfo {author} {\bibfnamefont {Q.}~\bibnamefont {Fan}}, \bibinfo {author} {\bibfnamefont {S.}~\bibnamefont {Guan}}, \bibinfo {author} {\bibfnamefont {G.}~\bibnamefont {Li}}, \bibinfo {author} {\bibfnamefont {P.}~\bibnamefont {Zhang}},\ and\ \bibinfo {author} {\bibfnamefont {T.}~\bibnamefont {Zhang}},\ }\bibfield  {title} {\bibinfo {title} {{Realization of strong coupling between deterministic single-atom arrays and a high-finesse miniature optical cavity}},\ }\href@noop {} {\bibfield  {journal} {\bibinfo  {journal} {Physical Review Letters}\ }\textbf {\bibinfo {volume} {130}},\ \bibinfo {pages} {173601} (\bibinfo {year} {2023})}\BibitemShut {NoStop}%
\bibitem [{\citenamefont {Grinkemeyer}\ \emph {et~al.}(2025)\citenamefont {Grinkemeyer}, \citenamefont {Guardado-Sanchez}, \citenamefont {Dimitrova}, \citenamefont {Shchepanovich}, \citenamefont {Mandopoulou}, \citenamefont {Borregaard}, \citenamefont {Vuleti{\'c}},\ and\ \citenamefont {Lukin}}]{grinkemeyer2025error}%
  \BibitemOpen
  \bibfield  {author} {\bibinfo {author} {\bibfnamefont {B.}~\bibnamefont {Grinkemeyer}}, \bibinfo {author} {\bibfnamefont {E.}~\bibnamefont {Guardado-Sanchez}}, \bibinfo {author} {\bibfnamefont {I.}~\bibnamefont {Dimitrova}}, \bibinfo {author} {\bibfnamefont {D.}~\bibnamefont {Shchepanovich}}, \bibinfo {author} {\bibfnamefont {G.~E.}\ \bibnamefont {Mandopoulou}}, \bibinfo {author} {\bibfnamefont {J.}~\bibnamefont {Borregaard}}, \bibinfo {author} {\bibfnamefont {V.}~\bibnamefont {Vuleti{\'c}}},\ and\ \bibinfo {author} {\bibfnamefont {M.~D.}\ \bibnamefont {Lukin}},\ }\bibfield  {title} {\bibinfo {title} {{Error-detected quantum operations with neutral atoms mediated by an optical cavity}},\ }\href@noop {} {\bibfield  {journal} {\bibinfo  {journal} {Science}\ }\textbf {\bibinfo {volume} {387}},\ \bibinfo {pages} {1301} (\bibinfo {year} {2025})}\BibitemShut {NoStop}%
\bibitem [{\citenamefont {Hartung}\ \emph {et~al.}(2024)\citenamefont {Hartung}, \citenamefont {Seubert}, \citenamefont {Welte}, \citenamefont {Distante},\ and\ \citenamefont {Rempe}}]{hartung2024quantum}%
  \BibitemOpen
  \bibfield  {author} {\bibinfo {author} {\bibfnamefont {L.}~\bibnamefont {Hartung}}, \bibinfo {author} {\bibfnamefont {M.}~\bibnamefont {Seubert}}, \bibinfo {author} {\bibfnamefont {S.}~\bibnamefont {Welte}}, \bibinfo {author} {\bibfnamefont {E.}~\bibnamefont {Distante}},\ and\ \bibinfo {author} {\bibfnamefont {G.}~\bibnamefont {Rempe}},\ }\bibfield  {title} {\bibinfo {title} {{A quantum-network register assembled with optical tweezers in an optical cavity}},\ }\href@noop {} {\bibfield  {journal} {\bibinfo  {journal} {Science}\ }\textbf {\bibinfo {volume} {385}},\ \bibinfo {pages} {179} (\bibinfo {year} {2024})}\BibitemShut {NoStop}%
\bibitem [{\citenamefont {Hu}\ \emph {et~al.}(2025)\citenamefont {Hu}, \citenamefont {Sinclair}, \citenamefont {Bytyqi}, \citenamefont {Chong}, \citenamefont {Rudelis}, \citenamefont {Ramette}, \citenamefont {Vendeiro},\ and\ \citenamefont {Vuleti{\'c}}}]{hu2025site}%
  \BibitemOpen
  \bibfield  {author} {\bibinfo {author} {\bibfnamefont {B.}~\bibnamefont {Hu}}, \bibinfo {author} {\bibfnamefont {J.}~\bibnamefont {Sinclair}}, \bibinfo {author} {\bibfnamefont {E.}~\bibnamefont {Bytyqi}}, \bibinfo {author} {\bibfnamefont {M.}~\bibnamefont {Chong}}, \bibinfo {author} {\bibfnamefont {A.}~\bibnamefont {Rudelis}}, \bibinfo {author} {\bibfnamefont {J.}~\bibnamefont {Ramette}}, \bibinfo {author} {\bibfnamefont {Z.}~\bibnamefont {Vendeiro}},\ and\ \bibinfo {author} {\bibfnamefont {V.}~\bibnamefont {Vuleti{\'c}}},\ }\bibfield  {title} {\bibinfo {title} {{Site-selective cavity readout and classical error correction of a 5-bit atomic register}},\ }\href@noop {} {\bibfield  {journal} {\bibinfo  {journal} {Physical Review Letters}\ }\textbf {\bibinfo {volume} {134}},\ \bibinfo {pages} {120801} (\bibinfo {year} {2025})}\BibitemShut {NoStop}%
\bibitem [{\citenamefont {Savary}\ and\ \citenamefont {Balents}(2016)}]{savary2016quantum}%
  \BibitemOpen
  \bibfield  {author} {\bibinfo {author} {\bibfnamefont {L.}~\bibnamefont {Savary}}\ and\ \bibinfo {author} {\bibfnamefont {L.}~\bibnamefont {Balents}},\ }\bibfield  {title} {\bibinfo {title} {{Quantum spin liquids: a review}},\ }\href@noop {} {\bibfield  {journal} {\bibinfo  {journal} {Reports on Progress in Physics}\ }\textbf {\bibinfo {volume} {80}},\ \bibinfo {pages} {016502} (\bibinfo {year} {2016})}\BibitemShut {NoStop}%
\bibitem [{\citenamefont {Dimer}\ \emph {et~al.}(2007)\citenamefont {Dimer}, \citenamefont {Estienne}, \citenamefont {Parkins},\ and\ \citenamefont {Carmichael}}]{dimer2007proposed}%
  \BibitemOpen
  \bibfield  {author} {\bibinfo {author} {\bibfnamefont {F.}~\bibnamefont {Dimer}}, \bibinfo {author} {\bibfnamefont {B.}~\bibnamefont {Estienne}}, \bibinfo {author} {\bibfnamefont {A.}~\bibnamefont {Parkins}},\ and\ \bibinfo {author} {\bibfnamefont {H.}~\bibnamefont {Carmichael}},\ }\bibfield  {title} {\bibinfo {title} {{Proposed realization of the Dicke-model quantum phase transition in an optical cavity QED system}},\ }\href@noop {} {\bibfield  {journal} {\bibinfo  {journal} {Physical Review A}\ }\textbf {\bibinfo {volume} {75}},\ \bibinfo {pages} {013804} (\bibinfo {year} {2007})}\BibitemShut {NoStop}%
\bibitem [{\citenamefont {Zeiher}\ \emph {et~al.}(2016)\citenamefont {Zeiher}, \citenamefont {Van~Bijnen}, \citenamefont {Schau{\ss}}, \citenamefont {Hild}, \citenamefont {Choi}, \citenamefont {Pohl}, \citenamefont {Bloch},\ and\ \citenamefont {Gross}}]{zeiher2016many}%
  \BibitemOpen
  \bibfield  {author} {\bibinfo {author} {\bibfnamefont {J.}~\bibnamefont {Zeiher}}, \bibinfo {author} {\bibfnamefont {R.}~\bibnamefont {Van~Bijnen}}, \bibinfo {author} {\bibfnamefont {P.}~\bibnamefont {Schau{\ss}}}, \bibinfo {author} {\bibfnamefont {S.}~\bibnamefont {Hild}}, \bibinfo {author} {\bibfnamefont {J.-y.}\ \bibnamefont {Choi}}, \bibinfo {author} {\bibfnamefont {T.}~\bibnamefont {Pohl}}, \bibinfo {author} {\bibfnamefont {I.}~\bibnamefont {Bloch}},\ and\ \bibinfo {author} {\bibfnamefont {C.}~\bibnamefont {Gross}},\ }\bibfield  {title} {\bibinfo {title} {{Many-body interferometry of a Rydberg-dressed spin lattice}},\ }\href@noop {} {\bibfield  {journal} {\bibinfo  {journal} {Nature Physics}\ }\textbf {\bibinfo {volume} {12}},\ \bibinfo {pages} {1095} (\bibinfo {year} {2016})}\BibitemShut {NoStop}%
\bibitem [{\citenamefont {Lee}\ and\ \citenamefont {Johnson}(2004)}]{lee2004first}%
  \BibitemOpen
  \bibfield  {author} {\bibinfo {author} {\bibfnamefont {C.~F.}\ \bibnamefont {Lee}}\ and\ \bibinfo {author} {\bibfnamefont {N.~F.}\ \bibnamefont {Johnson}},\ }\bibfield  {title} {\bibinfo {title} {{First-order superradiant phase transitions in a multiqubit cavity system}},\ }\href@noop {} {\bibfield  {journal} {\bibinfo  {journal} {Physical Review Letters}\ }\textbf {\bibinfo {volume} {93}},\ \bibinfo {pages} {083001} (\bibinfo {year} {2004})}\BibitemShut {NoStop}%
\bibitem [{\citenamefont {Gammelmark}\ and\ \citenamefont {M{\o}lmer}(2011)}]{gammelmark2011phase}%
  \BibitemOpen
  \bibfield  {author} {\bibinfo {author} {\bibfnamefont {S.}~\bibnamefont {Gammelmark}}\ and\ \bibinfo {author} {\bibfnamefont {K.}~\bibnamefont {M{\o}lmer}},\ }\bibfield  {title} {\bibinfo {title} {{Phase transitions and Heisenberg limited metrology in an Ising chain interacting with a single-mode cavity field}},\ }\href@noop {} {\bibfield  {journal} {\bibinfo  {journal} {New Journal of Physics}\ }\textbf {\bibinfo {volume} {13}},\ \bibinfo {pages} {053035} (\bibinfo {year} {2011})}\BibitemShut {NoStop}%
\bibitem [{\citenamefont {Gammelmark}\ and\ \citenamefont {M{\o}lmer}(2012)}]{gammelmark2012interacting}%
  \BibitemOpen
  \bibfield  {author} {\bibinfo {author} {\bibfnamefont {S.}~\bibnamefont {Gammelmark}}\ and\ \bibinfo {author} {\bibfnamefont {K.}~\bibnamefont {M{\o}lmer}},\ }\bibfield  {title} {\bibinfo {title} {{Interacting spins in a cavity: finite-size effects and symmetry-breaking dynamics}},\ }\href@noop {} {\bibfield  {journal} {\bibinfo  {journal} {Physical Review A}\ }\textbf {\bibinfo {volume} {85}},\ \bibinfo {pages} {042114} (\bibinfo {year} {2012})}\BibitemShut {NoStop}%
\bibitem [{\citenamefont {Zhang}\ \emph {et~al.}(2013)\citenamefont {Zhang}, \citenamefont {Sun}, \citenamefont {Wen}, \citenamefont {Liu}, \citenamefont {Eggert},\ and\ \citenamefont {Ji}}]{zhang2013rydberg}%
  \BibitemOpen
  \bibfield  {author} {\bibinfo {author} {\bibfnamefont {X.-F.}\ \bibnamefont {Zhang}}, \bibinfo {author} {\bibfnamefont {Q.}~\bibnamefont {Sun}}, \bibinfo {author} {\bibfnamefont {Y.-C.}\ \bibnamefont {Wen}}, \bibinfo {author} {\bibfnamefont {W.-M.}\ \bibnamefont {Liu}}, \bibinfo {author} {\bibfnamefont {S.}~\bibnamefont {Eggert}},\ and\ \bibinfo {author} {\bibfnamefont {A.-C.}\ \bibnamefont {Ji}},\ }\bibfield  {title} {\bibinfo {title} {{Rydberg polaritons in a cavity: A superradiant solid}},\ }\href@noop {} {\bibfield  {journal} {\bibinfo  {journal} {Physical Review Letters}\ }\textbf {\bibinfo {volume} {110}},\ \bibinfo {pages} {090402} (\bibinfo {year} {2013})}\BibitemShut {NoStop}%
\bibitem [{\citenamefont {Zhang}\ \emph {et~al.}(2014)\citenamefont {Zhang}, \citenamefont {Yu}, \citenamefont {Liang}, \citenamefont {Chen}, \citenamefont {Jia},\ and\ \citenamefont {Nori}}]{zhang2014quantum}%
  \BibitemOpen
  \bibfield  {author} {\bibinfo {author} {\bibfnamefont {Y.}~\bibnamefont {Zhang}}, \bibinfo {author} {\bibfnamefont {L.}~\bibnamefont {Yu}}, \bibinfo {author} {\bibfnamefont {J.-Q.}\ \bibnamefont {Liang}}, \bibinfo {author} {\bibfnamefont {G.}~\bibnamefont {Chen}}, \bibinfo {author} {\bibfnamefont {S.}~\bibnamefont {Jia}},\ and\ \bibinfo {author} {\bibfnamefont {F.}~\bibnamefont {Nori}},\ }\bibfield  {title} {\bibinfo {title} {{Quantum phases in circuit QED with a superconducting qubit array}},\ }\href@noop {} {\bibfield  {journal} {\bibinfo  {journal} {Scientific Reports}\ }\textbf {\bibinfo {volume} {4}},\ \bibinfo {pages} {4083} (\bibinfo {year} {2014})}\BibitemShut {NoStop}%
\bibitem [{\citenamefont {Gelhausen}\ \emph {et~al.}(2016)\citenamefont {Gelhausen}, \citenamefont {Buchhold}, \citenamefont {Rosch},\ and\ \citenamefont {Strack}}]{gelhausen2016quantum}%
  \BibitemOpen
  \bibfield  {author} {\bibinfo {author} {\bibfnamefont {J.}~\bibnamefont {Gelhausen}}, \bibinfo {author} {\bibfnamefont {M.}~\bibnamefont {Buchhold}}, \bibinfo {author} {\bibfnamefont {A.}~\bibnamefont {Rosch}},\ and\ \bibinfo {author} {\bibfnamefont {P.}~\bibnamefont {Strack}},\ }\bibfield  {title} {\bibinfo {title} {{Quantum-optical magnets with competing short-and long-range interactions: Rydberg-dressed spin lattice in an optical cavity}},\ }\href@noop {} {\bibfield  {journal} {\bibinfo  {journal} {SciPost Physics}\ }\textbf {\bibinfo {volume} {1}},\ \bibinfo {pages} {004} (\bibinfo {year} {2016})}\BibitemShut {NoStop}%
\bibitem [{\citenamefont {Rohn}\ \emph {et~al.}(2020)\citenamefont {Rohn}, \citenamefont {H{\"o}rmann}, \citenamefont {Genes},\ and\ \citenamefont {Schmidt}}]{rohn2020ising}%
  \BibitemOpen
  \bibfield  {author} {\bibinfo {author} {\bibfnamefont {J.}~\bibnamefont {Rohn}}, \bibinfo {author} {\bibfnamefont {M.}~\bibnamefont {H{\"o}rmann}}, \bibinfo {author} {\bibfnamefont {C.}~\bibnamefont {Genes}},\ and\ \bibinfo {author} {\bibfnamefont {K.~P.}\ \bibnamefont {Schmidt}},\ }\bibfield  {title} {\bibinfo {title} {{Ising model in a light-induced quantized transverse field}},\ }\href@noop {} {\bibfield  {journal} {\bibinfo  {journal} {Physical Review Research}\ }\textbf {\bibinfo {volume} {2}},\ \bibinfo {pages} {023131} (\bibinfo {year} {2020})}\BibitemShut {NoStop}%
\bibitem [{\citenamefont {Schuler}\ \emph {et~al.}(2020)\citenamefont {Schuler}, \citenamefont {De~Bernardis}, \citenamefont {L{\"a}uchli},\ and\ \citenamefont {Rabl}}]{schuler2020vacua}%
  \BibitemOpen
  \bibfield  {author} {\bibinfo {author} {\bibfnamefont {M.}~\bibnamefont {Schuler}}, \bibinfo {author} {\bibfnamefont {D.}~\bibnamefont {De~Bernardis}}, \bibinfo {author} {\bibfnamefont {A.}~\bibnamefont {L{\"a}uchli}},\ and\ \bibinfo {author} {\bibfnamefont {P.}~\bibnamefont {Rabl}},\ }\bibfield  {title} {\bibinfo {title} {{The vacua of dipolar cavity quantum electrodynamics}},\ }\href@noop {} {\bibfield  {journal} {\bibinfo  {journal} {SciPost Physics}\ }\textbf {\bibinfo {volume} {9}},\ \bibinfo {pages} {066} (\bibinfo {year} {2020})}\BibitemShut {NoStop}%
\bibitem [{\citenamefont {An}\ \emph {et~al.}(2022)\citenamefont {An}, \citenamefont {Zhou}, \citenamefont {Wang},\ and\ \citenamefont {Zhang}}]{an2022quantum}%
  \BibitemOpen
  \bibfield  {author} {\bibinfo {author} {\bibfnamefont {G.-Q.}\ \bibnamefont {An}}, \bibinfo {author} {\bibfnamefont {Y.-H.}\ \bibnamefont {Zhou}}, \bibinfo {author} {\bibfnamefont {T.}~\bibnamefont {Wang}},\ and\ \bibinfo {author} {\bibfnamefont {X.-F.}\ \bibnamefont {Zhang}},\ }\bibfield  {title} {\bibinfo {title} {{Quantum phase transition of a two-dimensional Rydberg atom array in an optical cavity}},\ }\href@noop {} {\bibfield  {journal} {\bibinfo  {journal} {Physical Review B}\ }\textbf {\bibinfo {volume} {106}},\ \bibinfo {pages} {134506} (\bibinfo {year} {2022})}\BibitemShut {NoStop}%
\bibitem [{\citenamefont {Puel}\ and\ \citenamefont {Macr{\`\i}}(2024)}]{puel2024confined}%
  \BibitemOpen
  \bibfield  {author} {\bibinfo {author} {\bibfnamefont {T.~O.}\ \bibnamefont {Puel}}\ and\ \bibinfo {author} {\bibfnamefont {T.}~\bibnamefont {Macr{\`\i}}},\ }\bibfield  {title} {\bibinfo {title} {{Confined meson excitations in Rydberg-atom arrays coupled to a cavity field}},\ }\href@noop {} {\bibfield  {journal} {\bibinfo  {journal} {Physical Review Letters}\ }\textbf {\bibinfo {volume} {133}},\ \bibinfo {pages} {106901} (\bibinfo {year} {2024})}\BibitemShut {NoStop}%
\bibitem [{\citenamefont {Bacciconi}\ \emph {et~al.}(2025)\citenamefont {Bacciconi}, \citenamefont {Xavier}, \citenamefont {Marinelli}, \citenamefont {Bhakuni},\ and\ \citenamefont {Dalmonte}}]{bacciconi2025local}%
  \BibitemOpen
  \bibfield  {author} {\bibinfo {author} {\bibfnamefont {Z.}~\bibnamefont {Bacciconi}}, \bibinfo {author} {\bibfnamefont {H.~B.}\ \bibnamefont {Xavier}}, \bibinfo {author} {\bibfnamefont {M.}~\bibnamefont {Marinelli}}, \bibinfo {author} {\bibfnamefont {D.~S.}\ \bibnamefont {Bhakuni}},\ and\ \bibinfo {author} {\bibfnamefont {M.}~\bibnamefont {Dalmonte}},\ }\bibfield  {title} {\bibinfo {title} {{Local vs nonlocal dynamics in cavity-coupled Rydberg atom arrays}},\ }\href@noop {} {\bibfield  {journal} {\bibinfo  {journal} {Physical Review Letters}\ }\textbf {\bibinfo {volume} {134}},\ \bibinfo {pages} {213604} (\bibinfo {year} {2025})}\BibitemShut {NoStop}%
\bibitem [{\citenamefont {Chiocchetta}\ \emph {et~al.}(2021)\citenamefont {Chiocchetta}, \citenamefont {Kiese}, \citenamefont {Zelle}, \citenamefont {Piazza},\ and\ \citenamefont {Diehl}}]{chiocchetta2021cavity}%
  \BibitemOpen
  \bibfield  {author} {\bibinfo {author} {\bibfnamefont {A.}~\bibnamefont {Chiocchetta}}, \bibinfo {author} {\bibfnamefont {D.}~\bibnamefont {Kiese}}, \bibinfo {author} {\bibfnamefont {C.~P.}\ \bibnamefont {Zelle}}, \bibinfo {author} {\bibfnamefont {F.}~\bibnamefont {Piazza}},\ and\ \bibinfo {author} {\bibfnamefont {S.}~\bibnamefont {Diehl}},\ }\bibfield  {title} {\bibinfo {title} {{Cavity-induced quantum spin liquids}},\ }\href@noop {} {\bibfield  {journal} {\bibinfo  {journal} {Nature Communications}\ }\textbf {\bibinfo {volume} {12}},\ \bibinfo {pages} {5901} (\bibinfo {year} {2021})}\BibitemShut {NoStop}%
\bibitem [{\citenamefont {Marshall}(1955)}]{marshall1955antiferromagnetism}%
  \BibitemOpen
  \bibfield  {author} {\bibinfo {author} {\bibfnamefont {W.}~\bibnamefont {Marshall}},\ }\bibfield  {title} {\bibinfo {title} {{Antiferromagnetism}},\ }\href@noop {} {\bibfield  {journal} {\bibinfo  {journal} {Proceedings of the Royal Society of London. Series A}\ }\textbf {\bibinfo {volume} {232}},\ \bibinfo {pages} {48} (\bibinfo {year} {1955})}\BibitemShut {NoStop}%
\bibitem [{\citenamefont {Lieb}\ and\ \citenamefont {Mattis}(1962)}]{lieb1962ordering}%
  \BibitemOpen
  \bibfield  {author} {\bibinfo {author} {\bibfnamefont {E.}~\bibnamefont {Lieb}}\ and\ \bibinfo {author} {\bibfnamefont {D.}~\bibnamefont {Mattis}},\ }\bibfield  {title} {\bibinfo {title} {{Ordering energy levels of interacting spin systems}},\ }\href@noop {} {\bibfield  {journal} {\bibinfo  {journal} {Journal of Mathematical Physics}\ }\textbf {\bibinfo {volume} {3}},\ \bibinfo {pages} {749} (\bibinfo {year} {1962})}\BibitemShut {NoStop}%
\bibitem [{\citenamefont {Richter}\ and\ \citenamefont {Schulenburg}(2010)}]{richter2010spin}%
  \BibitemOpen
  \bibfield  {author} {\bibinfo {author} {\bibfnamefont {J.}~\bibnamefont {Richter}}\ and\ \bibinfo {author} {\bibfnamefont {J.}~\bibnamefont {Schulenburg}},\ }\bibfield  {title} {\bibinfo {title} {{The spin-$1/2$ $J_1$-$J_2$ Heisenberg antiferromagnet on the square lattice: Exact diagonalization for $N=40$ spins}},\ }\href@noop {} {\bibfield  {journal} {\bibinfo  {journal} {The European Physical Journal B}\ }\textbf {\bibinfo {volume} {73}},\ \bibinfo {pages} {117} (\bibinfo {year} {2010})}\BibitemShut {NoStop}%
\bibitem [{\citenamefont {Jiang}\ \emph {et~al.}(2012{\natexlab{a}})\citenamefont {Jiang}, \citenamefont {Yao},\ and\ \citenamefont {Balents}}]{jiang2012spin}%
  \BibitemOpen
  \bibfield  {author} {\bibinfo {author} {\bibfnamefont {H.-C.}\ \bibnamefont {Jiang}}, \bibinfo {author} {\bibfnamefont {H.}~\bibnamefont {Yao}},\ and\ \bibinfo {author} {\bibfnamefont {L.}~\bibnamefont {Balents}},\ }\bibfield  {title} {\bibinfo {title} {{Spin liquid ground state of the spin-$1/2$ square $J_1$-$J_2$ Heisenberg model}},\ }\href@noop {} {\bibfield  {journal} {\bibinfo  {journal} {Physical Review B}\ }\textbf {\bibinfo {volume} {86}},\ \bibinfo {pages} {024424} (\bibinfo {year} {2012}{\natexlab{a}})}\BibitemShut {NoStop}%
\bibitem [{\citenamefont {Hu}\ \emph {et~al.}(2013)\citenamefont {Hu}, \citenamefont {Becca}, \citenamefont {Parola},\ and\ \citenamefont {Sorella}}]{hu2013direct}%
  \BibitemOpen
  \bibfield  {author} {\bibinfo {author} {\bibfnamefont {W.-J.}\ \bibnamefont {Hu}}, \bibinfo {author} {\bibfnamefont {F.}~\bibnamefont {Becca}}, \bibinfo {author} {\bibfnamefont {A.}~\bibnamefont {Parola}},\ and\ \bibinfo {author} {\bibfnamefont {S.}~\bibnamefont {Sorella}},\ }\bibfield  {title} {\bibinfo {title} {{Direct evidence for a gapless $Z_2$ spin liquid by frustrating N\'eel antiferromagnetism}},\ }\href@noop {} {\bibfield  {journal} {\bibinfo  {journal} {Physical Review B}\ }\textbf {\bibinfo {volume} {88}},\ \bibinfo {pages} {060402} (\bibinfo {year} {2013})}\BibitemShut {NoStop}%
\bibitem [{\citenamefont {Wang}\ \emph {et~al.}(2013)\citenamefont {Wang}, \citenamefont {Poilblanc}, \citenamefont {Gu}, \citenamefont {Wen},\ and\ \citenamefont {Verstraete}}]{wang2013constructing}%
  \BibitemOpen
  \bibfield  {author} {\bibinfo {author} {\bibfnamefont {L.}~\bibnamefont {Wang}}, \bibinfo {author} {\bibfnamefont {D.}~\bibnamefont {Poilblanc}}, \bibinfo {author} {\bibfnamefont {Z.-C.}\ \bibnamefont {Gu}}, \bibinfo {author} {\bibfnamefont {X.-G.}\ \bibnamefont {Wen}},\ and\ \bibinfo {author} {\bibfnamefont {F.}~\bibnamefont {Verstraete}},\ }\bibfield  {title} {\bibinfo {title} {{Constructing a gapless spin-liquid state for the spin-$1/2$ $J_1$-$J_2$ Heisenberg model on a square lattice}},\ }\href@noop {} {\bibfield  {journal} {\bibinfo  {journal} {Physical Review Letters}\ }\textbf {\bibinfo {volume} {111}},\ \bibinfo {pages} {037202} (\bibinfo {year} {2013})}\BibitemShut {NoStop}%
\bibitem [{\citenamefont {Wang}\ and\ \citenamefont {Sandvik}(2018)}]{wang2018critical}%
  \BibitemOpen
  \bibfield  {author} {\bibinfo {author} {\bibfnamefont {L.}~\bibnamefont {Wang}}\ and\ \bibinfo {author} {\bibfnamefont {A.~W.}\ \bibnamefont {Sandvik}},\ }\bibfield  {title} {\bibinfo {title} {{Critical level crossings and gapless spin liquid in the square-lattice spin-$1/2$ $J_1$-$J_2$ Heisenberg antiferromagnet}},\ }\href@noop {} {\bibfield  {journal} {\bibinfo  {journal} {Physical Review Letters}\ }\textbf {\bibinfo {volume} {121}},\ \bibinfo {pages} {107202} (\bibinfo {year} {2018})}\BibitemShut {NoStop}%
\bibitem [{\citenamefont {Nomura}\ and\ \citenamefont {Imada}(2021)}]{nomura2021dirac}%
  \BibitemOpen
  \bibfield  {author} {\bibinfo {author} {\bibfnamefont {Y.}~\bibnamefont {Nomura}}\ and\ \bibinfo {author} {\bibfnamefont {M.}~\bibnamefont {Imada}},\ }\bibfield  {title} {\bibinfo {title} {{Dirac-type nodal spin liquid revealed by refined quantum many-body solver using neural-network wave function, correlation ratio, and level spectroscopy}},\ }\href@noop {} {\bibfield  {journal} {\bibinfo  {journal} {Physical Review X}\ }\textbf {\bibinfo {volume} {11}},\ \bibinfo {pages} {031034} (\bibinfo {year} {2021})}\BibitemShut {NoStop}%
\bibitem [{\citenamefont {Liu}\ \emph {et~al.}(2022)\citenamefont {Liu}, \citenamefont {Gong}, \citenamefont {Li}, \citenamefont {Poilblanc}, \citenamefont {Chen},\ and\ \citenamefont {Gu}}]{liu2022gapless}%
  \BibitemOpen
  \bibfield  {author} {\bibinfo {author} {\bibfnamefont {W.-Y.}\ \bibnamefont {Liu}}, \bibinfo {author} {\bibfnamefont {S.-S.}\ \bibnamefont {Gong}}, \bibinfo {author} {\bibfnamefont {Y.-B.}\ \bibnamefont {Li}}, \bibinfo {author} {\bibfnamefont {D.}~\bibnamefont {Poilblanc}}, \bibinfo {author} {\bibfnamefont {W.-Q.}\ \bibnamefont {Chen}},\ and\ \bibinfo {author} {\bibfnamefont {Z.-C.}\ \bibnamefont {Gu}},\ }\bibfield  {title} {\bibinfo {title} {{Gapless quantum spin liquid and global phase diagram of the spin-$1/2$ $J_1$-$J_2$ square antiferromagnetic Heisenberg model}},\ }\href@noop {} {\bibfield  {journal} {\bibinfo  {journal} {Science bulletin}\ }\textbf {\bibinfo {volume} {67}},\ \bibinfo {pages} {1034} (\bibinfo {year} {2022})}\BibitemShut {NoStop}%
\bibitem [{\citenamefont {Zhu}\ and\ \citenamefont {White}(2015)}]{zhu2015spin}%
  \BibitemOpen
  \bibfield  {author} {\bibinfo {author} {\bibfnamefont {Z.}~\bibnamefont {Zhu}}\ and\ \bibinfo {author} {\bibfnamefont {S.~R.}\ \bibnamefont {White}},\ }\bibfield  {title} {\bibinfo {title} {{Spin liquid phase of the $S=1/2$ $J_1$-$J_2$ Heisenberg model on the triangular lattice}},\ }\href@noop {} {\bibfield  {journal} {\bibinfo  {journal} {Physical Review B}\ }\textbf {\bibinfo {volume} {92}},\ \bibinfo {pages} {041105} (\bibinfo {year} {2015})}\BibitemShut {NoStop}%
\bibitem [{\citenamefont {Hu}\ \emph {et~al.}(2015)\citenamefont {Hu}, \citenamefont {Gong}, \citenamefont {Zhu},\ and\ \citenamefont {Sheng}}]{hu2015competing}%
  \BibitemOpen
  \bibfield  {author} {\bibinfo {author} {\bibfnamefont {W.-J.}\ \bibnamefont {Hu}}, \bibinfo {author} {\bibfnamefont {S.-S.}\ \bibnamefont {Gong}}, \bibinfo {author} {\bibfnamefont {W.}~\bibnamefont {Zhu}},\ and\ \bibinfo {author} {\bibfnamefont {D.}~\bibnamefont {Sheng}},\ }\bibfield  {title} {\bibinfo {title} {{Competing spin-liquid states in the spin-$1/2$ Heisenberg model on the triangular lattice}},\ }\href@noop {} {\bibfield  {journal} {\bibinfo  {journal} {Physical Review B}\ }\textbf {\bibinfo {volume} {92}},\ \bibinfo {pages} {140403} (\bibinfo {year} {2015})}\BibitemShut {NoStop}%
\bibitem [{\citenamefont {Iqbal}\ \emph {et~al.}(2016)\citenamefont {Iqbal}, \citenamefont {Hu}, \citenamefont {Thomale}, \citenamefont {Poilblanc},\ and\ \citenamefont {Becca}}]{iqbal2016spin}%
  \BibitemOpen
  \bibfield  {author} {\bibinfo {author} {\bibfnamefont {Y.}~\bibnamefont {Iqbal}}, \bibinfo {author} {\bibfnamefont {W.-J.}\ \bibnamefont {Hu}}, \bibinfo {author} {\bibfnamefont {R.}~\bibnamefont {Thomale}}, \bibinfo {author} {\bibfnamefont {D.}~\bibnamefont {Poilblanc}},\ and\ \bibinfo {author} {\bibfnamefont {F.}~\bibnamefont {Becca}},\ }\bibfield  {title} {\bibinfo {title} {{Spin liquid nature in the Heisenberg $J_1$-$J_2$ triangular antiferromagnet}},\ }\href@noop {} {\bibfield  {journal} {\bibinfo  {journal} {Physical Review B}\ }\textbf {\bibinfo {volume} {93}},\ \bibinfo {pages} {144411} (\bibinfo {year} {2016})}\BibitemShut {NoStop}%
\bibitem [{\citenamefont {Hu}\ \emph {et~al.}(2019)\citenamefont {Hu}, \citenamefont {Zhu}, \citenamefont {Eggert},\ and\ \citenamefont {He}}]{hu2019dirac}%
  \BibitemOpen
  \bibfield  {author} {\bibinfo {author} {\bibfnamefont {S.}~\bibnamefont {Hu}}, \bibinfo {author} {\bibfnamefont {W.}~\bibnamefont {Zhu}}, \bibinfo {author} {\bibfnamefont {S.}~\bibnamefont {Eggert}},\ and\ \bibinfo {author} {\bibfnamefont {Y.-C.}\ \bibnamefont {He}},\ }\bibfield  {title} {\bibinfo {title} {{Dirac spin liquid on the spin-$1/2$ triangular Heisenberg antiferromagnet}},\ }\href@noop {} {\bibfield  {journal} {\bibinfo  {journal} {Physical Review Letters}\ }\textbf {\bibinfo {volume} {123}},\ \bibinfo {pages} {207203} (\bibinfo {year} {2019})}\BibitemShut {NoStop}%
\bibitem [{\citenamefont {Wietek}\ \emph {et~al.}(2024)\citenamefont {Wietek}, \citenamefont {Capponi},\ and\ \citenamefont {L{\"a}uchli}}]{wietek2024quantum}%
  \BibitemOpen
  \bibfield  {author} {\bibinfo {author} {\bibfnamefont {A.}~\bibnamefont {Wietek}}, \bibinfo {author} {\bibfnamefont {S.}~\bibnamefont {Capponi}},\ and\ \bibinfo {author} {\bibfnamefont {A.~M.}\ \bibnamefont {L{\"a}uchli}},\ }\bibfield  {title} {\bibinfo {title} {{Quantum electrodynamics in 2+1 dimensions as the organizing principle of a triangular lattice antiferromagnet}},\ }\href@noop {} {\bibfield  {journal} {\bibinfo  {journal} {Physical Review X}\ }\textbf {\bibinfo {volume} {14}},\ \bibinfo {pages} {021010} (\bibinfo {year} {2024})}\BibitemShut {NoStop}%
\bibitem [{\citenamefont {Yan}\ \emph {et~al.}(2011)\citenamefont {Yan}, \citenamefont {Huse},\ and\ \citenamefont {White}}]{yan2011spin}%
  \BibitemOpen
  \bibfield  {author} {\bibinfo {author} {\bibfnamefont {S.}~\bibnamefont {Yan}}, \bibinfo {author} {\bibfnamefont {D.~A.}\ \bibnamefont {Huse}},\ and\ \bibinfo {author} {\bibfnamefont {S.~R.}\ \bibnamefont {White}},\ }\bibfield  {title} {\bibinfo {title} {{Spin-liquid ground state of the $S=1/2$ kagome Heisenberg antiferromagnet}},\ }\href@noop {} {\bibfield  {journal} {\bibinfo  {journal} {Science}\ }\textbf {\bibinfo {volume} {332}},\ \bibinfo {pages} {1173} (\bibinfo {year} {2011})}\BibitemShut {NoStop}%
\bibitem [{\citenamefont {Jiang}\ \emph {et~al.}(2012{\natexlab{b}})\citenamefont {Jiang}, \citenamefont {Wang},\ and\ \citenamefont {Balents}}]{jiang2012identifying}%
  \BibitemOpen
  \bibfield  {author} {\bibinfo {author} {\bibfnamefont {H.-C.}\ \bibnamefont {Jiang}}, \bibinfo {author} {\bibfnamefont {Z.}~\bibnamefont {Wang}},\ and\ \bibinfo {author} {\bibfnamefont {L.}~\bibnamefont {Balents}},\ }\bibfield  {title} {\bibinfo {title} {{Identifying topological order by entanglement entropy}},\ }\href@noop {} {\bibfield  {journal} {\bibinfo  {journal} {Nature Physics}\ }\textbf {\bibinfo {volume} {8}},\ \bibinfo {pages} {902} (\bibinfo {year} {2012}{\natexlab{b}})}\BibitemShut {NoStop}%
\bibitem [{\citenamefont {Depenbrock}\ \emph {et~al.}(2012)\citenamefont {Depenbrock}, \citenamefont {McCulloch},\ and\ \citenamefont {Schollw{\"o}ck}}]{depenbrock2012nature}%
  \BibitemOpen
  \bibfield  {author} {\bibinfo {author} {\bibfnamefont {S.}~\bibnamefont {Depenbrock}}, \bibinfo {author} {\bibfnamefont {I.~P.}\ \bibnamefont {McCulloch}},\ and\ \bibinfo {author} {\bibfnamefont {U.}~\bibnamefont {Schollw{\"o}ck}},\ }\bibfield  {title} {\bibinfo {title} {{Nature of the spin-liquid ground state of the $S=1/2$ Heisenberg model on the kagome lattice}},\ }\href@noop {} {\bibfield  {journal} {\bibinfo  {journal} {Physical Review Letters}\ }\textbf {\bibinfo {volume} {109}},\ \bibinfo {pages} {067201} (\bibinfo {year} {2012})}\BibitemShut {NoStop}%
\bibitem [{\citenamefont {Kolley}\ \emph {et~al.}(2015)\citenamefont {Kolley}, \citenamefont {Depenbrock}, \citenamefont {McCulloch}, \citenamefont {Schollw{\"o}ck},\ and\ \citenamefont {Alba}}]{kolley2015phase}%
  \BibitemOpen
  \bibfield  {author} {\bibinfo {author} {\bibfnamefont {F.}~\bibnamefont {Kolley}}, \bibinfo {author} {\bibfnamefont {S.}~\bibnamefont {Depenbrock}}, \bibinfo {author} {\bibfnamefont {I.~P.}\ \bibnamefont {McCulloch}}, \bibinfo {author} {\bibfnamefont {U.}~\bibnamefont {Schollw{\"o}ck}},\ and\ \bibinfo {author} {\bibfnamefont {V.}~\bibnamefont {Alba}},\ }\bibfield  {title} {\bibinfo {title} {{Phase diagram of the $J_1$-$J_2$ Heisenberg model on the kagome lattice}},\ }\href@noop {} {\bibfield  {journal} {\bibinfo  {journal} {Physical Review B}\ }\textbf {\bibinfo {volume} {91}},\ \bibinfo {pages} {104418} (\bibinfo {year} {2015})}\BibitemShut {NoStop}%
\bibitem [{\citenamefont {Iqbal}\ \emph {et~al.}(2013)\citenamefont {Iqbal}, \citenamefont {Becca}, \citenamefont {Sorella},\ and\ \citenamefont {Poilblanc}}]{iqbal2013gapless}%
  \BibitemOpen
  \bibfield  {author} {\bibinfo {author} {\bibfnamefont {Y.}~\bibnamefont {Iqbal}}, \bibinfo {author} {\bibfnamefont {F.}~\bibnamefont {Becca}}, \bibinfo {author} {\bibfnamefont {S.}~\bibnamefont {Sorella}},\ and\ \bibinfo {author} {\bibfnamefont {D.}~\bibnamefont {Poilblanc}},\ }\bibfield  {title} {\bibinfo {title} {{Gapless spin-liquid phase in the kagome spin-$1/2$ Heisenberg antiferromagnet}},\ }\href@noop {} {\bibfield  {journal} {\bibinfo  {journal} {Physical Review B}\ }\textbf {\bibinfo {volume} {87}},\ \bibinfo {pages} {060405} (\bibinfo {year} {2013})}\BibitemShut {NoStop}%
\bibitem [{\citenamefont {Liao}\ \emph {et~al.}(2017)\citenamefont {Liao}, \citenamefont {Xie}, \citenamefont {Chen}, \citenamefont {Liu}, \citenamefont {Xie}, \citenamefont {Huang}, \citenamefont {Normand},\ and\ \citenamefont {Xiang}}]{liao2017gapless}%
  \BibitemOpen
  \bibfield  {author} {\bibinfo {author} {\bibfnamefont {H.-J.}\ \bibnamefont {Liao}}, \bibinfo {author} {\bibfnamefont {Z.-Y.}\ \bibnamefont {Xie}}, \bibinfo {author} {\bibfnamefont {J.}~\bibnamefont {Chen}}, \bibinfo {author} {\bibfnamefont {Z.-Y.}\ \bibnamefont {Liu}}, \bibinfo {author} {\bibfnamefont {H.-D.}\ \bibnamefont {Xie}}, \bibinfo {author} {\bibfnamefont {R.-Z.}\ \bibnamefont {Huang}}, \bibinfo {author} {\bibfnamefont {B.}~\bibnamefont {Normand}},\ and\ \bibinfo {author} {\bibfnamefont {T.}~\bibnamefont {Xiang}},\ }\bibfield  {title} {\bibinfo {title} {{Gapless spin-liquid ground state in the $S=1/2$ kagome antiferromagnet}},\ }\href@noop {} {\bibfield  {journal} {\bibinfo  {journal} {Physical Review Letters}\ }\textbf {\bibinfo {volume} {118}},\ \bibinfo {pages} {137202} (\bibinfo {year} {2017})}\BibitemShut {NoStop}%
\bibitem [{\citenamefont {He}\ \emph {et~al.}(2017)\citenamefont {He}, \citenamefont {Zaletel}, \citenamefont {Oshikawa},\ and\ \citenamefont {Pollmann}}]{he2017signatures}%
  \BibitemOpen
  \bibfield  {author} {\bibinfo {author} {\bibfnamefont {Y.-C.}\ \bibnamefont {He}}, \bibinfo {author} {\bibfnamefont {M.~P.}\ \bibnamefont {Zaletel}}, \bibinfo {author} {\bibfnamefont {M.}~\bibnamefont {Oshikawa}},\ and\ \bibinfo {author} {\bibfnamefont {F.}~\bibnamefont {Pollmann}},\ }\bibfield  {title} {\bibinfo {title} {{Signatures of Dirac cones in a DMRG study of the kagome Heisenberg model}},\ }\href@noop {} {\bibfield  {journal} {\bibinfo  {journal} {Physical Review X}\ }\textbf {\bibinfo {volume} {7}},\ \bibinfo {pages} {031020} (\bibinfo {year} {2017})}\BibitemShut {NoStop}%
\bibitem [{\citenamefont {Reuther}\ \emph {et~al.}(2014)\citenamefont {Reuther}, \citenamefont {Lee},\ and\ \citenamefont {Alicea}}]{reuther2014classification}%
  \BibitemOpen
  \bibfield  {author} {\bibinfo {author} {\bibfnamefont {J.}~\bibnamefont {Reuther}}, \bibinfo {author} {\bibfnamefont {S.-P.}\ \bibnamefont {Lee}},\ and\ \bibinfo {author} {\bibfnamefont {J.}~\bibnamefont {Alicea}},\ }\bibfield  {title} {\bibinfo {title} {{Classification of spin liquids on the square lattice with strong spin-orbit coupling}},\ }\href@noop {} {\bibfield  {journal} {\bibinfo  {journal} {Physical Review B}\ }\textbf {\bibinfo {volume} {90}},\ \bibinfo {pages} {174417} (\bibinfo {year} {2014})}\BibitemShut {NoStop}%
\bibitem [{\citenamefont {Block}\ \emph {et~al.}(2024)\citenamefont {Block}, \citenamefont {Ye}, \citenamefont {Roberts}, \citenamefont {Chern}, \citenamefont {Wu}, \citenamefont {Wang}, \citenamefont {Pollet}, \citenamefont {Davis}, \citenamefont {Halperin},\ and\ \citenamefont {Yao}}]{block2024scalable}%
  \BibitemOpen
  \bibfield  {author} {\bibinfo {author} {\bibfnamefont {M.}~\bibnamefont {Block}}, \bibinfo {author} {\bibfnamefont {B.}~\bibnamefont {Ye}}, \bibinfo {author} {\bibfnamefont {B.}~\bibnamefont {Roberts}}, \bibinfo {author} {\bibfnamefont {S.}~\bibnamefont {Chern}}, \bibinfo {author} {\bibfnamefont {W.}~\bibnamefont {Wu}}, \bibinfo {author} {\bibfnamefont {Z.}~\bibnamefont {Wang}}, \bibinfo {author} {\bibfnamefont {L.}~\bibnamefont {Pollet}}, \bibinfo {author} {\bibfnamefont {E.~J.}\ \bibnamefont {Davis}}, \bibinfo {author} {\bibfnamefont {B.~I.}\ \bibnamefont {Halperin}},\ and\ \bibinfo {author} {\bibfnamefont {N.~Y.}\ \bibnamefont {Yao}},\ }\bibfield  {title} {\bibinfo {title} {{Scalable spin squeezing from finite-temperature easy-plane magnetism}},\ }\href@noop {} {\bibfield  {journal} {\bibinfo  {journal} {Nature Physics}\ }\textbf {\bibinfo {volume} {20}},\ \bibinfo {pages} {1575} (\bibinfo {year} {2024})}\BibitemShut {NoStop}%
\bibitem [{\citenamefont {Kaubruegger}\ \emph {et~al.}(2025)\citenamefont {Kaubruegger}, \citenamefont {Padilla}, \citenamefont {Shankar}, \citenamefont {Hotter}, \citenamefont {Muleady}, \citenamefont {Bringewatt}, \citenamefont {Baamara}, \citenamefont {Abbasgholinejad}, \citenamefont {Gorshkov}, \citenamefont {M{\o}lmer} \emph {et~al.}}]{kaubruegger2025lieb}%
  \BibitemOpen
  \bibfield  {author} {\bibinfo {author} {\bibfnamefont {R.}~\bibnamefont {Kaubruegger}}, \bibinfo {author} {\bibfnamefont {D.~F.}\ \bibnamefont {Padilla}}, \bibinfo {author} {\bibfnamefont {A.}~\bibnamefont {Shankar}}, \bibinfo {author} {\bibfnamefont {C.}~\bibnamefont {Hotter}}, \bibinfo {author} {\bibfnamefont {S.~R.}\ \bibnamefont {Muleady}}, \bibinfo {author} {\bibfnamefont {J.}~\bibnamefont {Bringewatt}}, \bibinfo {author} {\bibfnamefont {Y.}~\bibnamefont {Baamara}}, \bibinfo {author} {\bibfnamefont {E.}~\bibnamefont {Abbasgholinejad}}, \bibinfo {author} {\bibfnamefont {A.~V.}\ \bibnamefont {Gorshkov}}, \bibinfo {author} {\bibfnamefont {K.}~\bibnamefont {M{\o}lmer}}, \emph {et~al.},\ }\bibfield  {title} {\bibinfo {title} {{Lieb-Mattis states for robust entangled differential phase sensing}},\ }\href@noop {} {\bibfield  {journal} {\bibinfo  {journal} {arXiv preprint arXiv:2506.10151}\ } (\bibinfo {year} {2025})}\BibitemShut {NoStop}%
\bibitem [{\citenamefont {Bravyi}\ \emph {et~al.}(2011)\citenamefont {Bravyi}, \citenamefont {DiVincenzo},\ and\ \citenamefont {Loss}}]{bravyi2011schrieffer}%
  \BibitemOpen
  \bibfield  {author} {\bibinfo {author} {\bibfnamefont {S.}~\bibnamefont {Bravyi}}, \bibinfo {author} {\bibfnamefont {D.~P.}\ \bibnamefont {DiVincenzo}},\ and\ \bibinfo {author} {\bibfnamefont {D.}~\bibnamefont {Loss}},\ }\bibfield  {title} {\bibinfo {title} {{Schrieffer--Wolff transformation for quantum many-body systems}},\ }\href@noop {} {\bibfield  {journal} {\bibinfo  {journal} {Annals of physics}\ }\textbf {\bibinfo {volume} {326}},\ \bibinfo {pages} {2793} (\bibinfo {year} {2011})}\BibitemShut {NoStop}%
\bibitem [{\citenamefont {Cable}\ and\ \citenamefont {Durkin}(2010)}]{cable2010parameter}%
  \BibitemOpen
  \bibfield  {author} {\bibinfo {author} {\bibfnamefont {H.}~\bibnamefont {Cable}}\ and\ \bibinfo {author} {\bibfnamefont {G.~A.}\ \bibnamefont {Durkin}},\ }\bibfield  {title} {\bibinfo {title} {{Parameter estimation with entangled photons produced by parametric down-conversion}},\ }\href@noop {} {\bibfield  {journal} {\bibinfo  {journal} {Physical Review Letters}\ }\textbf {\bibinfo {volume} {105}},\ \bibinfo {pages} {013603} (\bibinfo {year} {2010})}\BibitemShut {NoStop}%
\bibitem [{\citenamefont {Kitzinger}\ \emph {et~al.}(2020)\citenamefont {Kitzinger}, \citenamefont {Chaudhary}, \citenamefont {Kondappan}, \citenamefont {Ivannikov},\ and\ \citenamefont {Byrnes}}]{kitzinger2020two}%
  \BibitemOpen
  \bibfield  {author} {\bibinfo {author} {\bibfnamefont {J.}~\bibnamefont {Kitzinger}}, \bibinfo {author} {\bibfnamefont {M.}~\bibnamefont {Chaudhary}}, \bibinfo {author} {\bibfnamefont {M.}~\bibnamefont {Kondappan}}, \bibinfo {author} {\bibfnamefont {V.}~\bibnamefont {Ivannikov}},\ and\ \bibinfo {author} {\bibfnamefont {T.}~\bibnamefont {Byrnes}},\ }\bibfield  {title} {\bibinfo {title} {{Two-axis two-spin squeezed states}},\ }\href@noop {} {\bibfield  {journal} {\bibinfo  {journal} {Physical Review Research}\ }\textbf {\bibinfo {volume} {2}},\ \bibinfo {pages} {033504} (\bibinfo {year} {2020})}\BibitemShut {NoStop}%
\bibitem [{\citenamefont {Sundar}\ \emph {et~al.}(2023)\citenamefont {Sundar}, \citenamefont {Barberena}, \citenamefont {Orioli}, \citenamefont {Chu}, \citenamefont {Thompson}, \citenamefont {Rey},\ and\ \citenamefont {Lewis-Swan}}]{sundar2023bosonic}%
  \BibitemOpen
  \bibfield  {author} {\bibinfo {author} {\bibfnamefont {B.}~\bibnamefont {Sundar}}, \bibinfo {author} {\bibfnamefont {D.}~\bibnamefont {Barberena}}, \bibinfo {author} {\bibfnamefont {A.~P.}\ \bibnamefont {Orioli}}, \bibinfo {author} {\bibfnamefont {A.}~\bibnamefont {Chu}}, \bibinfo {author} {\bibfnamefont {J.~K.}\ \bibnamefont {Thompson}}, \bibinfo {author} {\bibfnamefont {A.~M.}\ \bibnamefont {Rey}},\ and\ \bibinfo {author} {\bibfnamefont {R.~J.}\ \bibnamefont {Lewis-Swan}},\ }\bibfield  {title} {\bibinfo {title} {{Bosonic pair production and squeezing for optical phase measurements in long-lived dipoles coupled to a cavity}},\ }\href@noop {} {\bibfield  {journal} {\bibinfo  {journal} {Physical Review Letters}\ }\textbf {\bibinfo {volume} {130}},\ \bibinfo {pages} {113202} (\bibinfo {year} {2023})}\BibitemShut {NoStop}%
\bibitem [{\citenamefont {Mamaev}\ \emph {et~al.}(2025)\citenamefont {Mamaev}, \citenamefont {Koppenh{\"o}fer}, \citenamefont {Pocklington},\ and\ \citenamefont {Clerk}}]{mamaev2025non}%
  \BibitemOpen
  \bibfield  {author} {\bibinfo {author} {\bibfnamefont {M.}~\bibnamefont {Mamaev}}, \bibinfo {author} {\bibfnamefont {M.}~\bibnamefont {Koppenh{\"o}fer}}, \bibinfo {author} {\bibfnamefont {A.}~\bibnamefont {Pocklington}},\ and\ \bibinfo {author} {\bibfnamefont {A.~A.}\ \bibnamefont {Clerk}},\ }\bibfield  {title} {\bibinfo {title} {{Non-Gaussian generalized two-mode squeezing: applications to two-ensemble spin squeezing and beyond}},\ }\href@noop {} {\bibfield  {journal} {\bibinfo  {journal} {Physical Review Letters}\ }\textbf {\bibinfo {volume} {134}},\ \bibinfo {pages} {073603} (\bibinfo {year} {2025})}\BibitemShut {NoStop}%
\bibitem [{\citenamefont {You}\ \emph {et~al.}(2007)\citenamefont {You}, \citenamefont {Li},\ and\ \citenamefont {Gu}}]{you2007fidelity}%
  \BibitemOpen
  \bibfield  {author} {\bibinfo {author} {\bibfnamefont {W.-L.}\ \bibnamefont {You}}, \bibinfo {author} {\bibfnamefont {Y.-W.}\ \bibnamefont {Li}},\ and\ \bibinfo {author} {\bibfnamefont {S.-J.}\ \bibnamefont {Gu}},\ }\bibfield  {title} {\bibinfo {title} {{Fidelity, dynamic structure factor, and susceptibility in critical phenomena}},\ }\href@noop {} {\bibfield  {journal} {\bibinfo  {journal} {Physical Review E}\ }\textbf {\bibinfo {volume} {76}},\ \bibinfo {pages} {022101} (\bibinfo {year} {2007})}\BibitemShut {NoStop}%
\bibitem [{\citenamefont {Gallegos}\ \emph {et~al.}(2025)\citenamefont {Gallegos}, \citenamefont {Jiang}, \citenamefont {White},\ and\ \citenamefont {Chernyshev}}]{gallegos2025phase}%
  \BibitemOpen
  \bibfield  {author} {\bibinfo {author} {\bibfnamefont {C.~A.}\ \bibnamefont {Gallegos}}, \bibinfo {author} {\bibfnamefont {S.}~\bibnamefont {Jiang}}, \bibinfo {author} {\bibfnamefont {S.~R.}\ \bibnamefont {White}},\ and\ \bibinfo {author} {\bibfnamefont {A.}~\bibnamefont {Chernyshev}},\ }\bibfield  {title} {\bibinfo {title} {{Phase diagram of the easy-axis triangular-lattice $J_1$-$J_2$ model}},\ }\href@noop {} {\bibfield  {journal} {\bibinfo  {journal} {Physical Review Letters}\ }\textbf {\bibinfo {volume} {134}},\ \bibinfo {pages} {196702} (\bibinfo {year} {2025})}\BibitemShut {NoStop}%
\bibitem [{\citenamefont {Colbois}\ \emph {et~al.}(2022)\citenamefont {Colbois}, \citenamefont {Vanhecke}, \citenamefont {Vanderstraeten}, \citenamefont {Smerald}, \citenamefont {Verstraete},\ and\ \citenamefont {Mila}}]{colbois2022partial}%
  \BibitemOpen
  \bibfield  {author} {\bibinfo {author} {\bibfnamefont {J.}~\bibnamefont {Colbois}}, \bibinfo {author} {\bibfnamefont {B.}~\bibnamefont {Vanhecke}}, \bibinfo {author} {\bibfnamefont {L.}~\bibnamefont {Vanderstraeten}}, \bibinfo {author} {\bibfnamefont {A.}~\bibnamefont {Smerald}}, \bibinfo {author} {\bibfnamefont {F.}~\bibnamefont {Verstraete}},\ and\ \bibinfo {author} {\bibfnamefont {F.}~\bibnamefont {Mila}},\ }\bibfield  {title} {\bibinfo {title} {{Partial lifting of degeneracy in the $J_1$-$J_2$-$J_3$ Ising antiferromagnet on the kagome lattice}},\ }\href@noop {} {\bibfield  {journal} {\bibinfo  {journal} {Physical Review B}\ }\textbf {\bibinfo {volume} {106}},\ \bibinfo {pages} {174403} (\bibinfo {year} {2022})}\BibitemShut {NoStop}%
\bibitem [{\citenamefont {He}\ and\ \citenamefont {Chen}(2015)}]{he2015distinct}%
  \BibitemOpen
  \bibfield  {author} {\bibinfo {author} {\bibfnamefont {Y.-C.}\ \bibnamefont {He}}\ and\ \bibinfo {author} {\bibfnamefont {Y.}~\bibnamefont {Chen}},\ }\bibfield  {title} {\bibinfo {title} {{Distinct spin liquids and their transitions in spin-$1/2$ XXZ kagome antiferromagnets}},\ }\href@noop {} {\bibfield  {journal} {\bibinfo  {journal} {Physical Review Letters}\ }\textbf {\bibinfo {volume} {114}},\ \bibinfo {pages} {037201} (\bibinfo {year} {2015})}\BibitemShut {NoStop}%
\bibitem [{\citenamefont {Lieb}\ and\ \citenamefont {Robinson}(1972)}]{lieb1972finite}%
  \BibitemOpen
  \bibfield  {author} {\bibinfo {author} {\bibfnamefont {E.~H.}\ \bibnamefont {Lieb}}\ and\ \bibinfo {author} {\bibfnamefont {D.~W.}\ \bibnamefont {Robinson}},\ }\bibfield  {title} {\bibinfo {title} {{The finite group velocity of quantum spin systems}},\ }\href@noop {} {\bibfield  {journal} {\bibinfo  {journal} {Communications in Mathematical Physics}\ }\textbf {\bibinfo {volume} {28}},\ \bibinfo {pages} {251} (\bibinfo {year} {1972})}\BibitemShut {NoStop}%
\bibitem [{\citenamefont {Bravyi}\ \emph {et~al.}(2010)\citenamefont {Bravyi}, \citenamefont {Hastings},\ and\ \citenamefont {Michalakis}}]{bravyi2010topological}%
  \BibitemOpen
  \bibfield  {author} {\bibinfo {author} {\bibfnamefont {S.}~\bibnamefont {Bravyi}}, \bibinfo {author} {\bibfnamefont {M.~B.}\ \bibnamefont {Hastings}},\ and\ \bibinfo {author} {\bibfnamefont {S.}~\bibnamefont {Michalakis}},\ }\bibfield  {title} {\bibinfo {title} {{Topological quantum order: stability under local perturbations}},\ }\href@noop {} {\bibfield  {journal} {\bibinfo  {journal} {Journal of Mathematical Physics}\ }\textbf {\bibinfo {volume} {51}} (\bibinfo {year} {2010})}\BibitemShut {NoStop}%
\bibitem [{\citenamefont {Gong}\ \emph {et~al.}(2015)\citenamefont {Gong}, \citenamefont {Zhu}, \citenamefont {Balents},\ and\ \citenamefont {Sheng}}]{gong2015global}%
  \BibitemOpen
  \bibfield  {author} {\bibinfo {author} {\bibfnamefont {S.-S.}\ \bibnamefont {Gong}}, \bibinfo {author} {\bibfnamefont {W.}~\bibnamefont {Zhu}}, \bibinfo {author} {\bibfnamefont {L.}~\bibnamefont {Balents}},\ and\ \bibinfo {author} {\bibfnamefont {D.}~\bibnamefont {Sheng}},\ }\bibfield  {title} {\bibinfo {title} {{Global phase diagram of competing ordered and quantum spin-liquid phases on the kagome lattice}},\ }\href@noop {} {\bibfield  {journal} {\bibinfo  {journal} {Physical Review B}\ }\textbf {\bibinfo {volume} {91}},\ \bibinfo {pages} {075112} (\bibinfo {year} {2015})}\BibitemShut {NoStop}%
\bibitem [{\citenamefont {Gong}\ \emph {et~al.}(2014)\citenamefont {Gong}, \citenamefont {Zhu},\ and\ \citenamefont {Sheng}}]{gong2014emergent}%
  \BibitemOpen
  \bibfield  {author} {\bibinfo {author} {\bibfnamefont {S.-S.}\ \bibnamefont {Gong}}, \bibinfo {author} {\bibfnamefont {W.}~\bibnamefont {Zhu}},\ and\ \bibinfo {author} {\bibfnamefont {D.}~\bibnamefont {Sheng}},\ }\bibfield  {title} {\bibinfo {title} {{Emergent chiral spin liquid: fractional quantum Hall effect in a kagome Heisenberg model}},\ }\href@noop {} {\bibfield  {journal} {\bibinfo  {journal} {Scientific Reports}\ }\textbf {\bibinfo {volume} {4}},\ \bibinfo {pages} {6317} (\bibinfo {year} {2014})}\BibitemShut {NoStop}%
\bibitem [{\citenamefont {Klich}(2010)}]{klich2010stability}%
  \BibitemOpen
  \bibfield  {author} {\bibinfo {author} {\bibfnamefont {I.}~\bibnamefont {Klich}},\ }\bibfield  {title} {\bibinfo {title} {{On the stability of topological phases on a lattice}},\ }\href@noop {} {\bibfield  {journal} {\bibinfo  {journal} {Annals of Physics}\ }\textbf {\bibinfo {volume} {325}},\ \bibinfo {pages} {2120} (\bibinfo {year} {2010})}\BibitemShut {NoStop}%
\bibitem [{\citenamefont {Hastings}\ and\ \citenamefont {Wen}(2005)}]{hastings2005quasiadiabatic}%
  \BibitemOpen
  \bibfield  {author} {\bibinfo {author} {\bibfnamefont {M.~B.}\ \bibnamefont {Hastings}}\ and\ \bibinfo {author} {\bibfnamefont {X.-G.}\ \bibnamefont {Wen}},\ }\bibfield  {title} {\bibinfo {title} {{Quasiadiabatic continuation of quantum states: The stability of topological ground-state degeneracy and emergent gauge invariance}},\ }\href@noop {} {\bibfield  {journal} {\bibinfo  {journal} {Physical Review B}\ }\textbf {\bibinfo {volume} {72}},\ \bibinfo {pages} {045141} (\bibinfo {year} {2005})}\BibitemShut {NoStop}%
\bibitem [{\citenamefont {Lapa}\ and\ \citenamefont {Levin}(2023)}]{lapa2023stability}%
  \BibitemOpen
  \bibfield  {author} {\bibinfo {author} {\bibfnamefont {M.~F.}\ \bibnamefont {Lapa}}\ and\ \bibinfo {author} {\bibfnamefont {M.}~\bibnamefont {Levin}},\ }\bibfield  {title} {\bibinfo {title} {{Stability of ground state degeneracy to long-range interactions}},\ }\href@noop {} {\bibfield  {journal} {\bibinfo  {journal} {Journal of Statistical Mechanics: Theory and Experiment}\ }\textbf {\bibinfo {volume} {2023}},\ \bibinfo {pages} {013102} (\bibinfo {year} {2023})}\BibitemShut {NoStop}%
\bibitem [{\citenamefont {Wietek}\ and\ \citenamefont {L{\"a}uchli}(2018)}]{wietek2018sublattice}%
  \BibitemOpen
  \bibfield  {author} {\bibinfo {author} {\bibfnamefont {A.}~\bibnamefont {Wietek}}\ and\ \bibinfo {author} {\bibfnamefont {A.~M.}\ \bibnamefont {L{\"a}uchli}},\ }\bibfield  {title} {\bibinfo {title} {{Sublattice coding algorithm and distributed memory parallelization for large-scale exact diagonalizations of quantum many-body systems}},\ }\href@noop {} {\bibfield  {journal} {\bibinfo  {journal} {Physical Review E}\ }\textbf {\bibinfo {volume} {98}},\ \bibinfo {pages} {033309} (\bibinfo {year} {2018})}\BibitemShut {NoStop}%
\bibitem [{\citenamefont {Wietek}\ \emph {et~al.}(2025)\citenamefont {Wietek}, \citenamefont {Staszewski}, \citenamefont {Ulaga}, \citenamefont {Ebert}, \citenamefont {Karlsson}, \citenamefont {Sarkar}, \citenamefont {Shackleton}, \citenamefont {Sinha},\ and\ \citenamefont {Soares}}]{wietek2025xdiag}%
  \BibitemOpen
  \bibfield  {author} {\bibinfo {author} {\bibfnamefont {A.}~\bibnamefont {Wietek}}, \bibinfo {author} {\bibfnamefont {L.}~\bibnamefont {Staszewski}}, \bibinfo {author} {\bibfnamefont {M.}~\bibnamefont {Ulaga}}, \bibinfo {author} {\bibfnamefont {P.~L.}\ \bibnamefont {Ebert}}, \bibinfo {author} {\bibfnamefont {H.}~\bibnamefont {Karlsson}}, \bibinfo {author} {\bibfnamefont {S.}~\bibnamefont {Sarkar}}, \bibinfo {author} {\bibfnamefont {H.}~\bibnamefont {Shackleton}}, \bibinfo {author} {\bibfnamefont {A.}~\bibnamefont {Sinha}},\ and\ \bibinfo {author} {\bibfnamefont {R.~D.}\ \bibnamefont {Soares}},\ }\bibfield  {title} {\bibinfo {title} {{XDiag: exact diagonalization for quantum many-body systems}},\ }\href@noop {} {\bibfield  {journal} {\bibinfo  {journal} {arXiv preprint arXiv:2505.02901}\ } (\bibinfo {year} {2025})}\BibitemShut {NoStop}%
\end{thebibliography}%

\newpage

\section*{Methods}

\subsection*{Variational energy of the squeezed AFM ansatz}

\noindent The squeezed AFM ansatz in Eq.\,\eqref{eq:SqueezedAFM} is constructed from the sublattice Dicke states
%----------------------
\begin{equation}
    \ket{m}_{\xi}
    = \frac{(S_+^{\,\xi})^{s+m}}{\sqrt{(s+m)!\,(s-m)!}}\,|\!\downarrow^{\otimes 2s}\rangle_\xi\,,
\end{equation}
%----------------------
where $\xi=A,B$ labels the two sublattices which contain $N_A=N_B=N/2$ sites. Therefore, the ansatz is invariant under  independent permutations of sites within each sublattice, $U(\pi_A,\pi_B)\ket{\theta}=\ket{\theta}$, where $U(\pi_A,\pi_B)=U(\pi_A)\otimes U(\pi_B)$ and $U(\pi_\xi)$ denotes the unitary operator implementing a permutation $\pi_\xi $ of sites within sublattice $\xi$. To evaluate the variational energy, it is convenient to symmetrize the Hamiltonian over all permutations
%----------------------
\begin{equation}
     \tilde{H}=\frac{1}{N_A ! N_B !}\sum_{\pi_A,\pi_B} U(\pi_A,\pi_B) H U^\dagger(\pi_A,\pi_B) \,,
\end{equation}
%----------------------
which preserves expectation value, $\langle H \rangle_\theta=\langle \tilde{H} \rangle_\theta$. The symmetrized Hamiltonian can then be expressed in terms of collective sublattice spin operators 
%----------------------
\begin{equation}
     \tilde{H}=\ \lambda \left[(S_x^A+S_x^B)^2 +(S_y^A+S_y^B)^2)\right] + \frac{8J}{N} S_A^z S_B^z\,.
\end{equation}
%----------------------
For brevity, we have omitted a term proportional to $(S_z^A + S_z^B)^2$ since it vanishes identically for the squeezed AFM ansatz due to its zero net magnetization. The variational energy evaluates to Eq.\,\eqref{eq:Variational_Energy} in the main text, and the relevant structure factors are given by
%----------------------
\begin{align}
\begin{split}
    \mathcal{S}_{xx}(\mathbf{0})= \frac{\mathcal{A}^2}{N} \sum_{n=0}^{2s}(\tanh\theta)^{2n} \Big\{ [s(s+1)-(s-n)^2] \\
    -  \tanh\theta [s(s+1)-(s-n)(s-n-1)] \Big\} 
\end{split}\label{eq:Sxx0}\\
\begin{split}
    \mathcal{S}_{xx}(\mathbf{Q})= \frac{\mathcal{A}^2}{N} \sum_{n=0}^{2s}(\tanh\theta)^{2n} \Big\{ [s(s+1)-(s-n)^2] \\
    +  \tanh\theta [s(s+1)-(s-n)(s-n-1)] \Big\} 
\end{split} \label{eq:SxxQ}\\
\begin{split}
    \mathcal{S}_{zz}(\mathbf{Q})=\frac{4\mathcal{A}^2}{N} \sum_{n=0}^{2s}(\tanh\theta)^{2n} (s-n)^2 
\end{split}\label{eq:SzzQ}
\end{align}
%---------------------- 
where $\mathcal{A}=\{\cosh^2\theta[1-(\tanh\theta)^{4s+2}] \}^{-1/2}$ is the normalization factor. The optimal squeezing parameter $\theta^\star$ can be found by numerically minimizing the variational energy $E(\theta^\star)=\displaystyle \min_\theta\{E(\theta)\}$.

\subsection*{Dyson-Maleev transformation}

\noindent To gain analytical insight, we can treat the sublattices as effective large spins and apply a Dyson–Maleev transformation
%----------------------
\begin{align}
&S^{A}_z=s-a^\dagger a     &  & S^{B}_z= b^\dagger b-s           \\
&S^{A}_+=\sqrt{2s}\Big(1-\frac{a^\dagger a}{2s}\Big)a         &  &    S^{B}_+= \sqrt{2s}b^\dagger\Big(1-\frac{b^\dagger b}{2s}\Big) \\
& S^{A}_-=\sqrt{2s}a^\dagger   &  & S^{B}_-=\sqrt{2s} b     
\end{align}
%----------------------
Here, $a^\dagger$ and $b^\dagger$ effectively create symmetric spin-wave excitations on the $A$ and $B$ sublattices, respectively, relative to the classical AFM state which is identified with the bosonic vacuum $\ket{s}_{\mathrm{A}}\ket{-s}_{\mathrm{B}} \to \ket{0}_{\mathrm{A}}\ket{0}_{\mathrm{B}} $. Formally, a faithful mapping requires a hard constraint that no more than $2s$ bosons occupy each sublattice Hilbert space, to match the dimension of the physical spin Hilbert space. However, for analytic tractability, this constraint is typically relaxed and therefore the mapping remains valid as long as the spin‑wave densities are relatively small $\langle a^\dagger a\rangle, \langle b^\dagger b\rangle \lesssim 2s$. 

In the bosonic language, the squeezed AFM state maps onto the two-mode squeezed vacuum state
%----------------------
\begin{equation}
    \ket{\theta}\to\mathrm{e}^{\theta(a^\dagger b^\dagger-ab)}\ket{0}_A\ket{0}_B\,.
    \label{eq:SqueezedAFM_Boson}
\end{equation}
%----------------------
This is a simple Gaussian state characterized by the following non-vanishing second-order correlations $ \langle a^\dagger a\rangle_\theta=\langle b^\dagger b\rangle_\theta= \sinh^2\theta$ and $\langle a^\dagger b^\dagger \rangle_\theta=\langle ab \rangle_\theta=-\frac{1}{2}\sinh 2\theta$,
while all higher-order correlations can be decomposed using Wick's theorem, e.g. $\langle a^\dagger b^\dagger a  b \rangle=\langle a^\dagger b^\dagger \rangle\langle a  b \rangle +\langle a^\dagger a \rangle\langle b^\dagger  b \rangle  $. This leads to the following expressions for the structure factors
%----------------------
\begin{align}
    &\mathcal{S}_{xx}(\mathbf{0})= \frac{1}{4} \mathrm{e}^{-2\theta} \left(1-\frac{4}{N}\sinh^2\theta\right)\\
    &\mathcal{S}_{xx}(\mathbf{Q})= \frac{1}{4} \mathrm{e}^{2\theta} \left(1-\frac{4}{N}\sinh^2\theta\right)\\
    &\mathcal{S}_{zz}(\mathbf{Q})=\frac{N}{4} - 2\sinh^2\theta \left(1-\frac{2}{N}\cosh 2\theta\right) 
\end{align}
%---------------------- 
where the squeezed and anti-squeezed correlations are now more explicit.

As expected, the bosonic variational energy approximates the full variational energy well in the weak-squeezing regime near the optimal squeezing point (Extended Data Fig.\,\ref{fig:Variational}\textcolor{RoyalBlue}{a}). As $\bar{\lambda}$ increases, the discrepancies increase and eventually the bosonic approximation breaks down, yielding no local minimum (Extended Data Fig.\,\ref{fig:Variational}\textcolor{RoyalBlue}{b}). To obtain an asymptotic expression for the optimal squeezing in the large-$N$ limit, we first simplify the variational energy by neglecting sub-leading terms in $N$
%----------------------
\begin{equation}
     E(\theta)  \sim -\frac{1}{2}JN+\frac{1}{2}\lambda N\mathrm{e}^{-2\theta}+J\mathrm{e}^{2\theta}-\frac{J}{N}\mathrm{e}^{4\theta}-\frac{1}{2}\lambda +2J\,,
\end{equation}
%---------------------
and determine the local minimum by solving the resulting depressed cubic equation. This yields the asymptotic expressions presented in the main text, where 
%---------------------
\begin{equation}
    \beta = \frac{2}{\sqrt{3}} \cos\left[ \frac{1}{3} \left( \frac{\pi}{2} + \arctan \left( \frac{\sqrt{27\bar{\lambda}/2J}}{\sqrt{1 - 27\bar{\lambda}/2J}} \right) \right) \right]
\,.
\end{equation}
%---------------------
A real-valued local minimum exists only when $27 \bar{\lambda} / 2J < 1$, and beyond this critical point the bosonic approximation breaks down. Within the regime of validity, the asymptotic energy expression in Eq.\,\eqref{eq:Min_Variational_Energy} agrees closely with the full numerical solution in the large-$N$ limit (Extended Data Fig.\,\ref{fig:Variational}\textcolor{RoyalBlue}{c}).

\subsection*{Ground state splitting in the weak-cavity limit}

\noindent Away from the maximally frustrated points, the classical AFM ground states of the $J_1$–$J_2$ Ising model on the square and triangular lattices have only a finite degeneracy, separated from the excited states by a non-zero energy gap. These degenerate configurations are related by $Z_2$ spin inversion and lattice rotations. Concretely, the square Néel AFM has a two-fold degeneracy, the square stripe AFM has a four-fold degeneracy, and the triangular stripe AFM has a six-fold degeneracy. This exact degeneracy is, however, accidental: any symmetry-preserving perturbation will generically lift it, and the true eigenstates will transform according to irreducible representations of the full symmetry group. Importantly, the resulting ground-state splitting typically decreases exponentially with system size, while the many-body excitation gap remains non-zero. In the thermodynamic limit, this leads to the spontaneous symmetry-breaking of the $Z_2$ spin inversion symmetry (and relevant lattice symmetries).

This behavior can be understood intuitively from a simple perturbative argument. For example, under a weak transverse field, the two AFM states related by spin inversion are macroscopically distinct, and therefore only couple at $N^{\mathrm{th}}$ order in perturbation theory. The resulting tunneling amplitude is exponentially small in system size, yielding an exponentially small ground-state splitting. These ideas have been placed on rigorous footing using cluster-expansion techniques \cite{klich2010stability} and quasi-adiabatic continuation arguments that rely on the existence of Lieb–Robinson bounds \cite{hastings2005quasiadiabatic}.

Although such techniques cannot be directly applied to systems with strong long-range interactions, recent work has extended these proofs to certain long-range models using a refined cluster expansion \cite{lapa2023stability}. A key condition of that analysis, however, is that the perturbation has an extensive operator norm, which is violated in the present case: in fact, with the rescaled interaction the norm of the cavity perturbation scales super-extensively as $\sim N^3$. Consequently, a different approach is required to determine whether a quasi-degenerate ground-state manifold persists with the cavity-mediated interactions.

To address this question, we perform variational analysis with the symmetrized squeezed AFM states
%----------------------
\begin{equation}
    \ket{\theta,\pm} =\frac{c_\pm}{\sqrt{2}} (\ket{\theta,1}\pm\ket{\theta,2})\,,
    \label{eq:SqueezedAFM_Sym}
\end{equation}
%----------------------
where $\ket{\theta,1}$ is given by Eq.\,\eqref{eq:SqueezedAFM} and $\ket{\theta,2}=\Pi\ket{\theta,1}$ is related by global spin inversion $\Pi=\prod_i \sigma_x^i$. Note that the normalization factor is non-trivial due to the non-orthogonality of the symmetry-broken states
%----------------------
\begin{equation}
    c_\pm = \big[1\pm \mathcal{A}^2(2s+1)(-\tanh\theta)^{2s} \big]^{-1/2}\,.
\end{equation}
%----------------------
The variational energies of these symmetrized states have the same form
%----------------------
\begin{equation}
    E_{\pm}(\theta)  = \lambda N \big[ \mathcal{S}_{xx}^{\pm}(\mathbf{0}) + \mathcal{S}^{\pm}_{yy}(\mathbf{0})\big] -2J\mathcal{S}_{zz}^{\pm}(\mathbf{Q})\,,
\end{equation}
%----------------------
where the corresponding structure factors can be written as
%----------------------
\begin{equation}
    \mathcal{S}_{\alpha\alpha}^{\pm}(\mathbf{q})=c_\pm^2 [\mathcal{S}_{\alpha\alpha}^{11}(\mathbf{q})\pm \mathcal{S}_{\alpha\alpha}^{12}(\mathbf{q})]\,,
\end{equation}
%----------------------
where
%----------------------
\begin{equation}
    \mathcal{S}_{\alpha\alpha}^{\mu\nu}(\mathbf{q})=\frac{1}{N}\sum_{ij}\mathrm{e}^{\mathrm{i} \mathbf{q}\cdot(\mathbf{r}_i-\mathbf{r}_j)} \bra{\theta,\mu} S_\alpha^i S_\alpha^j\ket{\theta,\nu}\,,
\end{equation}
%----------------------
are the matrix elements between the symmetry-broken states. The diagonal elements at the relevant momenta are the same as given in  Eqs.\,\eqref{eq:Sxx0}-\eqref{eq:SzzQ}, while the relevant off-diagonal elements are given by
%----------------------
\begin{align}
   &\mathcal{S}_{xx}^{12}(\mathbf{0}) = 
   \frac{\mathcal{A}^2}{N} f(s)(-\tanh\theta)^{2s}  (2-\tanh\theta - \coth\theta)\\
   &\mathcal{S}_{xx}^{12}(\mathbf{0}) = 
   \frac{\mathcal{A}^2}{N} f(s)(-\tanh\theta)^{2s}  (2+\tanh\theta + \coth\theta) \\
   &\mathcal{S}_{zz}^{12}(\mathbf{Q}) =
   \frac{4\mathcal{A}^2}{N} f(s)(-\tanh\theta)^{2s} 
\end{align}
%----------------------
where $f(s)=(2/3)s^3+s^2+(1/3)s$. As expected, the symmetrized ansatz always attains the lower energy compared to the symmetry-broken ansatz, with the improvement being marginal at weak squeezing (Extended Data Fig.\,\ref{fig:Variational}\textcolor{RoyalBlue}{a}) and progressively larger at stronger squeezing (Extended Data Fig.\,\ref{fig:Variational}\textcolor{RoyalBlue}{b}). 

The quantity of interest is the energy gap between the symmetric states, defined as $\Delta = \big|E_+(\theta^\star) - E_-(\theta^\star)\big|$. In the strong-squeezing limit, these states asymptotically approach different total-spin sectors, and thus we expect the gap to scale extensively $\Delta \sim N$. To gain analytical insight in the opposite limit of weak squeezing, we substitute the asymptotic form of the optimal squeezing parameter, which yields
%----------------------
\begin{equation}
   \Delta\sim 4JN  \mathrm{e}^{-\frac{1}{\sqrt{\eta} }}  \Bigg[ (1-\sqrt{\eta})(1+\beta^2)-\frac{1}{3\sqrt{\eta}}(1-\beta^2) \Bigg] \,,
   \label{eq:Gap}
\end{equation}
%----------------------
after neglecting sub-leading terms. This analysis reveals that the gap still scales extensively $\Delta \sim N$, but with a prefactor that is exponentially suppressed in the weak-cavity limit. We have verified that this analytical expression converges to the exact variational gap obtained numerically in the large-$N$ limit (Extended Data Fig.\,\ref{fig:Variational}\textcolor{RoyalBlue}{d}).

\subsection*{Exact diagonalization studies}

\noindent Numerically verifying the phase diagrams of the TCI model presents substantial challenges beyond those encountered in short-range models. The cavity-mediated all-to-all connectivity introduces entanglement patterns that lack a characteristic length scale, as exemplified by the squeezed AFM states. As a result, even in gapped phases, the ground state is not guaranteed to obey an area law for entanglement entropy. This undermines the efficiency of the matrix product state ansatz, which rely on area-law scaling to compress quantum states efficiently. In such long-range interacting systems, the required bond dimension may grow polynomially---or even exponentially---with system size. Additionally, DRMG algorithms perform local optimizations that are well-suited for short-range Hamiltonians. In contrast, in long-range systems like the TCI model, local updates may produce non-perturbative changes in the energy landscape, potentially leading to convergence issues. 

We will explore this further in future work, but here we focus on ED simulations, which avoids these issues by providing unbiased access to the many-body spectrum. However, ED is limited by the exponential growth of the Hilbert space which restricts us to small system sizes. In the TCI model, this challenge is compounded by the presence of long-range interactions: unlike short-range models where the Hamiltonian contains only $\mathcal{O}(N)$ terms, the TCI Hamiltonian includes $\mathcal{O}(N^2)$ interaction terms, yielding a much denser matrix. This significantly increases both memory requirements (if the matrix is stored explicitly) and the computational cost of matrix-vector multiplications.

To reach the largest possible system sizes, we fully exploit all available space-group symmetries, enabling block diagonalization of the Hamiltonian within different irreducible representations (irreps). Each irrep is denoted by the label $\mathbf{q}\cdot \chi_{\mathbf{q}}$, where $\mathbf{q}$ labels the Bloch momentum within the first Brillouin zone and $\chi_{\mathbf{q}}$ labels the irrep of the corresponding little point group. The symmetry-adapted bases are constructed using the efficient sublattice encoding algorithm \cite{wietek2018sublattice}, as part of the state-of-the-art XDiag.jl library \cite{wietek2025xdiag}, which we integrate with ARPACK.jl for iterative diagonalization using the implicitly restarted Lanczos method. 

For the spectra presented in Figs.\,\ref{fig:Heisenberg_Mapping}\textcolor{RoyalBlue}{a-c} we performed ED simulations on 24-site clusters with periodic boundary conditions, defined by the torus vectors (expressed in units of the primitive lattice vectors):
%----------------------
\begin{align*}
   \text{Square}&: \mathbf{T}_1=(6,0)\,,\, \mathbf{T}_2=(0,4) \\ 
   \text{Triangular}&: \mathbf{T}_1=(4,1)\,,\, \mathbf{T}_2=(-4,5) \\ 
   \text{Kagome}&: \mathbf{T}_1=(1,2)\,,\, \mathbf{T}_2=(-3,2) 
\end{align*}
%----------------------
We converged the lowest 20 eigenvalues of the TCI model in a few representative irreps, along with the corresponding lowest 20 singlet eigenvalues of the Heisenberg model. 

For the ground-state observables shown in Extended Data Figs.\,\ref{fig:ED_Square}-\ref{fig:ED_Kagome}, we performed ED calculations on larger 36-site clusters defined by the torus vectors:
%----------------------
\begin{align*}
   \text{Square}&: \mathbf{T}_1=(6,0)\,,\, \mathbf{T}_2=(0,6) \\ 
   \text{Triangular}&: \mathbf{T}_1=(6,0)\,,\, \mathbf{T}_2=(0,6) \\ 
   \text{Kagome}&: \mathbf{T}_1=(4,-2)\,,\, \mathbf{T}_2=(-2,4) 
\end{align*}
%----------------------
These high-symmetry clusters preserve the full point-group symmetry of the infinite lattice and accommodate the expected ordering wavevectors. Using a high-performance computer, we obtained the ground state in the zero-magnetization sector within the $\Gamma.A_1$ space-group irrep. Since the kagome lattice has fewer symmetries compared to the square and triangular lattices, were only able to perform calculations along the lines $J_2/J_1=0$ and $J_2/J_1=0.1$. The static spin structure factors were computed by summing over all symmetry-related momentum in the Brillouin zone.

\setcounter{figure}{0}
\renewcommand{\figurename}{Extended Data Fig.}
\renewcommand{\thefigure}{\arabic{figure}}
\makeatletter
\renewcommand{\theHfigure}{ED\arabic{figure}}
\makeatother

\newpage

%------------ ED FIGURE 1 -------------% 
\begin{figure*}[t]
\centering
\includegraphics[width=\textwidth]{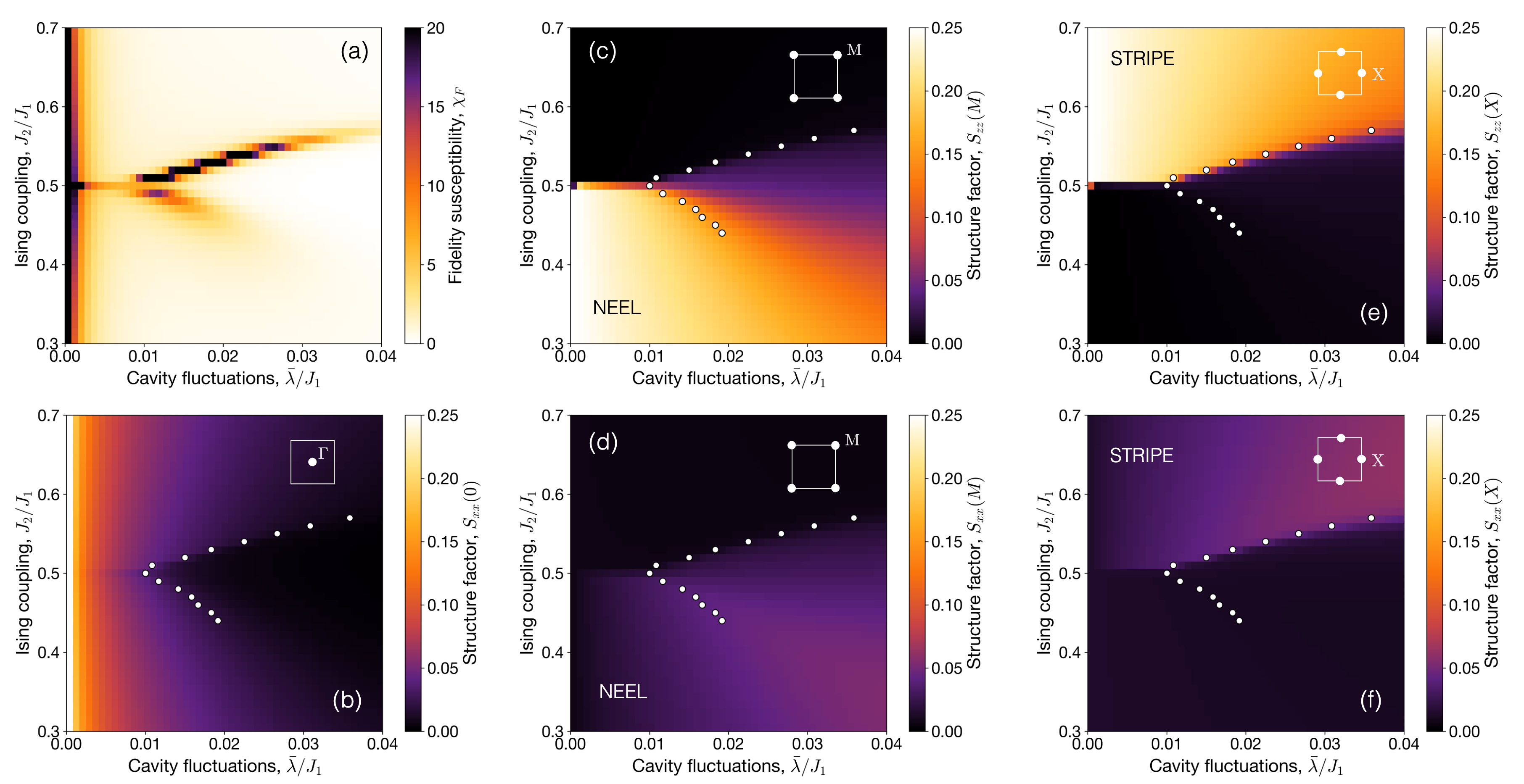}
\caption{\textbf{ED results for the $J_1$–$J_2$ TCI model on the square lattice.} (a) Fidelity susceptibility and (b–f) static spin structure factors of the ground state of the TCI model on the square lattice with $N=36$ sites, shown as functions of $J_2/J_1$ and $\bar{\lambda}/J_1$. The Néel and stripe phases exhibit dominant correlations at the ordering wavevectors $\mathbf{Q}=\mathrm{M}$ and $\mathbf{Q}=\mathrm{X}$, respectively. White dots mark the local maxima of the fidelity susceptibility, indicating possible phase boundaries. All calculations were performed in the zero-magnetization sector and within the $\Gamma.A_1$ space-group irrep, and the static structure factors are summed over symmetry-equivalent momentum. }
\label{fig:ED_Square}
\end{figure*}
%------------ ED FIGURE 1 -------------%

%------------ ED FIGURE 2 -------------% 
\begin{figure*}[t]
\centering
\includegraphics[width=\textwidth]{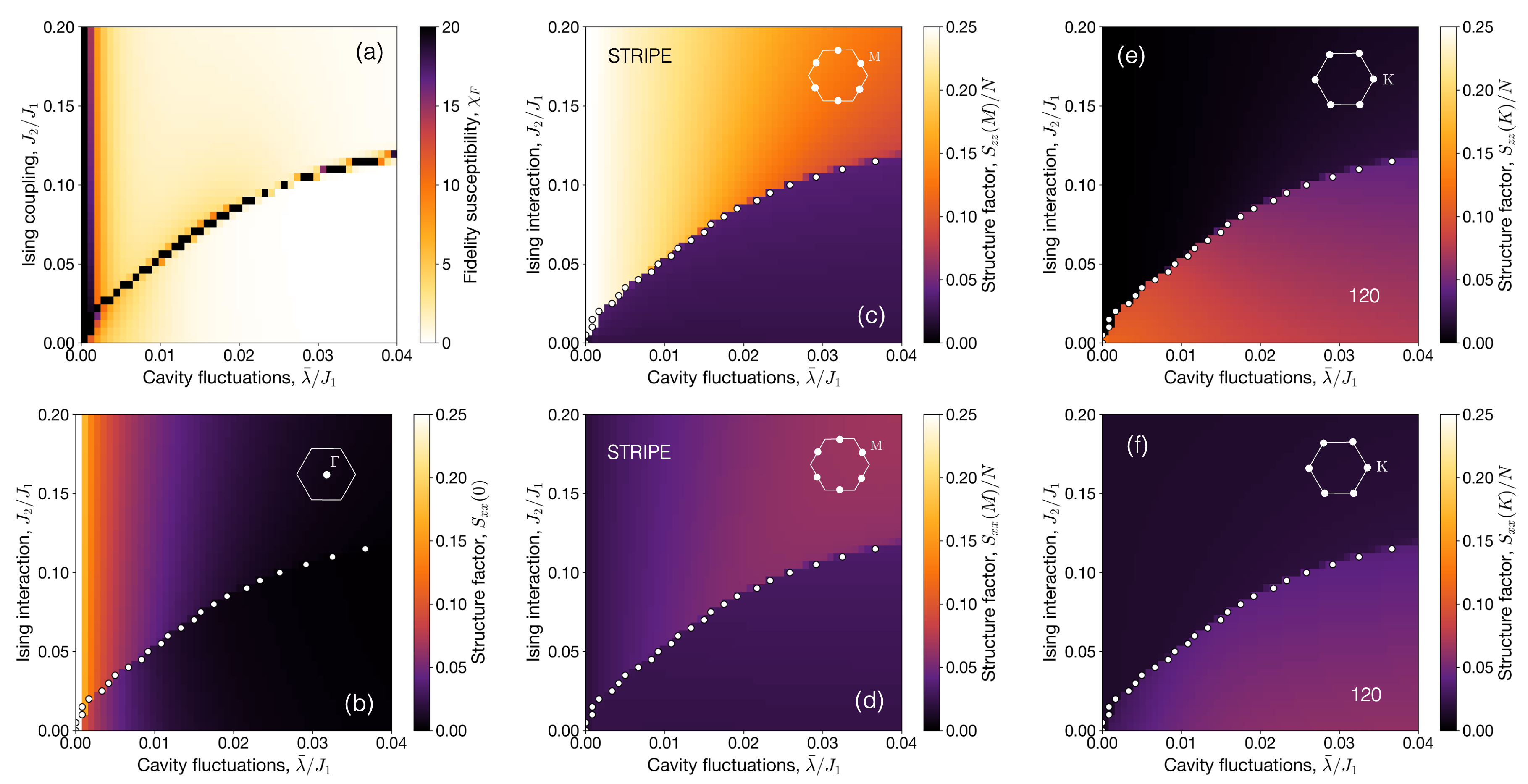}
\caption{\textbf{ED results for the $J_1$–$J_2$ TCI model on the triangular lattice.} (a) Fidelity susceptibility and (b–f) static spin structure factors of the ground state of the TCI model on the triangular lattice with $N=36$ sites, shown as functions of $J_2/J_1$ and $\bar{\lambda}/J_1$. The stripe and $120^\circ$ phases exhibit dominant correlations at the ordering wavevectors $\mathbf{Q}=\mathrm{M}$ and $\mathbf{Q}=\mathrm{K}$, respectively. White dots mark the local maxima of the fidelity susceptibility, indicating possible phase boundaries. All calculations were performed in the zero-magnetization sector and within the $\Gamma.A_1$ space-group irrep, and the static structure factors are summed over symmetry-equivalent momentum. }
\label{fig:ED_Triangular}
\end{figure*}
%------------ ED FIGURE 2 -------------%

%------------ ED FIGURE 3 -------------% 
\begin{figure*}[t]
\centering
\includegraphics[width=\textwidth]{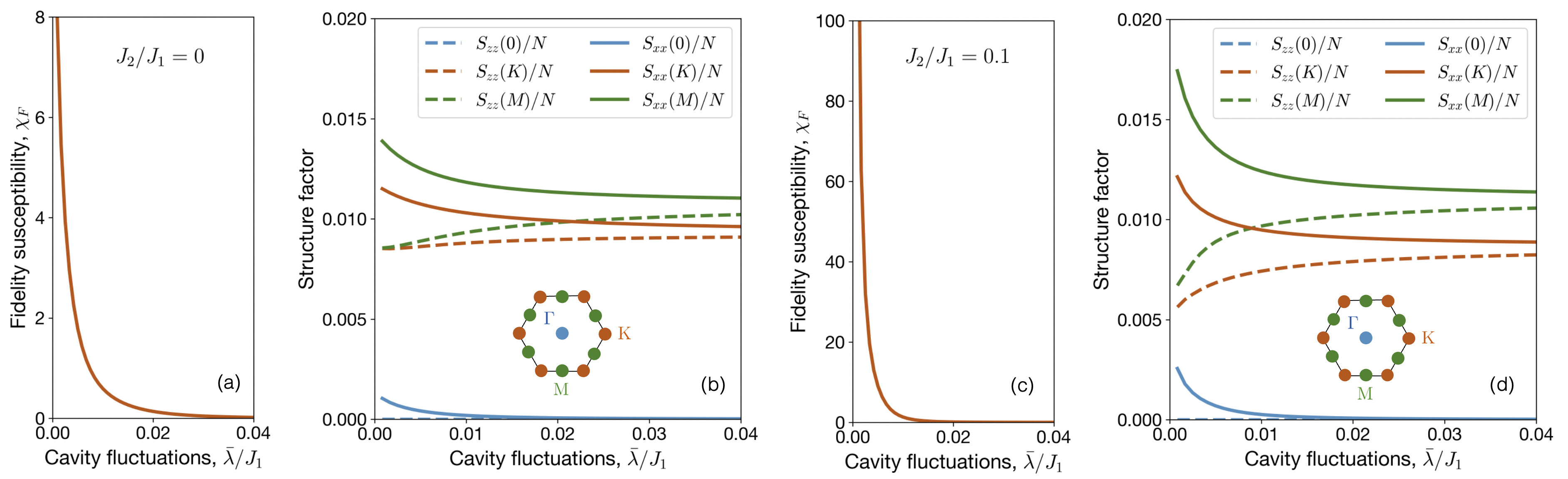}
\caption{\textbf{ED results for the $J_1$–$J_2$ TCI model on the kagome lattice.} (a) Fidelity susceptibility and (b) static spin structure factors of the ground state of the TCI model on the kagome lattice with $N=36$ sites, shown as function of $\bar{\lambda}/J_1$ for $J_2/J_1=0$. Panels (c) and (d) show the same quantities for $J_2/J_1=0.1$. All calculations were performed in the zero-magnetization sector and within the $\Gamma.A_1$ space-group irrep, and the static structure factors are summed over symmetry-equivalent momentum.}
\label{fig:ED_Kagome}
\end{figure*}
%------------ ED FIGURE 3 -------------%

%------------ ED FIGURE 4 -------------% 
\begin{figure*}[t]
\centering
\includegraphics[width=\textwidth]{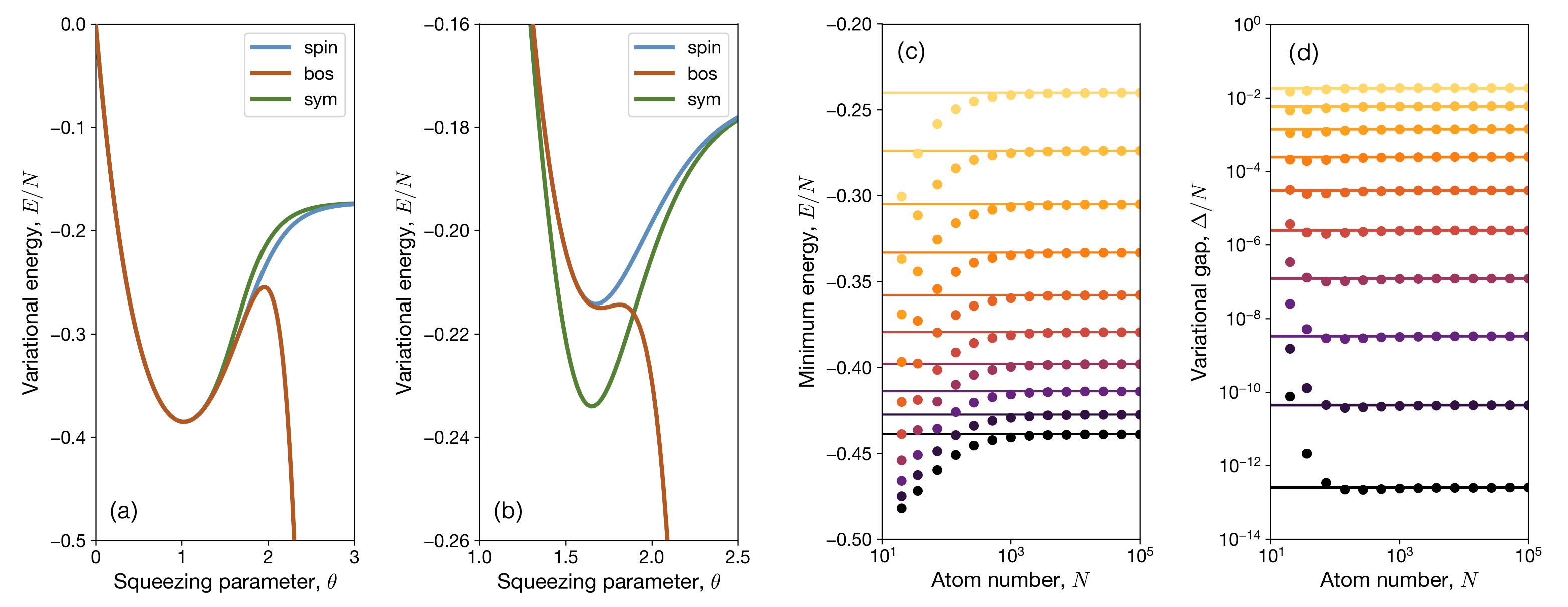}
\caption{\textbf{Squeezed AFM ansatz.}  (a-b) Variational energy of the squeezed AFM ansatz given by Eq.\,\eqref{eq:SqueezedAFM} (blue line), boson approximation given by Eq.\,\eqref{eq:SqueezedAFM_Boson}  (red line) and symmetrized ansatz given by Eq.\,\eqref{eq:SqueezedAFM_Sym} for $N=100$ with (a) $\bar{\lambda}/J_1=0.01$ and (b) $\bar{\lambda}/J_1=0.074$. (c) Minimum variational energy of the numerically optimized squeezed AFM state (dots) compared with the asymptotic expression given by Eq.\,\eqref{eq:Min_Variational_Energy} (lines). (d) Variational gap between the symmetrized squeezed AFM states obtained numerically (dots) compared with the asymptotic expression given by Eq.\,\eqref{eq:Gap} (lines). The colors correspond to logarithmically space values of $\bar{\lambda}/J_1$, ranging from $0.002$ (black) to $0.05$ (yellow). }
\label{fig:Variational}
\end{figure*}
%------------ ED FIGURE 4 -------------%

\end{document}